\documentclass[printer]{aa}
\usepackage{rotating}
\usepackage{graphicx}
\usepackage{txfonts}
\usepackage{natbib}
\usepackage[bookmarks=true,bookmarksopen=true,pdfstartview = FitV,pdfpagelayout = SinglePage,colorlinks = true,linkcolor=blue,citecolor=blue,urlcolor = blue]{hyperref} 
\newcommand{\sgra}{Sgr~A*}

\newcommand{\msun}{M_{\odot}}

\newcommand{\transquinze}{SWIFT~J174540.7-290015}

\begin{document}

\title{Continuation of the X-ray monitoring of \sgra{}: the increase in bright flaring rate confirmed}
\author{E.\ Mossoux\inst{1} \and B.\ Finociety\inst{2} \and J.-M.\ Beckers\inst{3} \and F.~H.\ Vincent\inst{4}}

\offprints{E.\ Mossoux}
\mail{@uliege.be}
\institute{Space sciences, Technologies and Astrophysics Research (STAR) Institute, Universit\'e de Li\`ege, All\'ee du 6 Ao\^ut, 19c, B\^at B5c, 4000 Li\`ege, Belgium
\and Universit\'e de Toulouse, IRAP, 14 avenue E.\ Belin, F-31400 Toulouse, France
\and GeoHydrodynamics and Environment Research (GHER), Universit\'e de Li\`ege, All\'ee du 6 Ao\^ut, 19c, B\^at B5a, 4000 Li\`ege, Belgium
\and LESIA, Observatoire de Paris, Université PSL, Sorbonne Universit\'e, Universit\'e Paris Diderot, Sorbonne Paris Cit\'e, 5 place Jules Janssen, 92195, Meudon, France}
\date{Received date / Accepted date}
\abstract{The supermassive black hole Sagittarius A* (\sgra{}) is located at the dynamical center of the Milky Way.
In a recent study of the X-ray flaring activity from \sgra{} using {\it Chandra}, {\it XMM-Newton,} and {\it Swift} observations from 1999 to 2015, it has been argued that the bright flaring rate has increased from 2014 August 31 while the faint flaring rate decreased from around 2013 August.}
{We tested the persistence of these changes in the
flaring rates with new X-ray observations of \sgra{} performed from 2016 to 2018 (total exposure of 1.4\,Ms).}
{We reprocessed the {\it Chandra}, {\it XMM-Newton,} and {\it Swift} observations from 2016 to 2018. 
We detected 9 flares in the {\it Chandra} data and 5 flares in the {\it Swift} data that we added to the set of 107 previously detected flares.
We computed the intrinsic distribution of flare fluxes and durations corrected for the sensitivity bias using a new method that allowed us to take the error on the flare fluxes and durations into account.
From this intrinsic distribution, we determined the average flare detection efficiency for each {\it Chandra}, {\it XMM-Newton,} and {\it Swift} observation. 
After correcting each observational exposure for this efficiency, we applied the Bayesian blocks algorithm on the concatenated flare arrival times.
As in the above-mentioned study, we also searched for a flux and fluence threshold that might lead to a change in flaring rate.
We improved the previous method by computing the average flare detection efficiencies for each flux and fluence range.}
{The Bayesian block algorithm did not detect any significant change in flaring rate of the 121 flares.
However, we detected an increase by a factor of about three in the flaring rate of the most luminous and most energetic flares that have occurred since 2014 August\ 30.}
{The X-ray activity of \sgra{} has increased for more than four years.
Additional studies about the overall near-infrared and radio behavior of \sgra{} are required to draw strong results on the multiwavelength activity of the black hole.}

\authorrunning{E. Mossoux et al.}
\titlerunning{Nineteen years of X-ray monitoring of Sgr~A*}
\maketitle

\section{Introduction}
The supermassive black hole (SMBH) Sagittarius~A* is located at the dynamical center of the Milky Way at a distance of about 8~kpc from the Sun \citep{genzel10, Falcke13}.
It has a low bolometric luminosity of about $10^{-9}$ times the Eddington luminosity of $3\times10^{44}\,\mathrm{erg\,s^{-1}}$ , computed for a mass of $4\times 10^6\,\msun{}$ \citep{schodel02, yuan03, ghez08, gillessen09}.
Several times a day, \sgra{} experiences increases in flux that are observable in radio (e.g., \citealt{zhao03, yusef-zadeh06c, yusef-zadeh08, yusef-zadeh09, marrone08}), near-infrared (NIR; e.g., \citealt{genzel03, yusef-zadeh06b, dodds-eden09, witzel12}), and X-rays (e.g., \citealt{baganoff01, porquet03, porquet08, neilsen13, barriere14}).
Several physical mechanisms have been studied to explain the origin of eruptions such as the accretion of additional fresh material onto the black hole \citep{yuan03, czerny13} or the change in distribution of the relativistic particles by the tidal disruption of asteroids passing close to the black hole \citep{cadez06, cadez08, kostic09, zubovas12}. 

\citet{neilsen13} performed the first statistical study of the X-ray flares of \sgra{}.
These authors analyzed the observations performed during the {\it Chandra} X-ray Visionary Project (XVP) in 2012.
They detected 39 X-ray flares that led to an observed X-ray flaring rate of $1.1\pm0.2\,$flares per day.
They detected flares using a Gaussian fitting on the light curves. 
However, this method is limited to flares with a duration longer than 400\,s and a peak count-rate higher than $0.015\,\mathrm{count\,s^{-1}}$ with the ACIS-S3 camera.
They also compared their Gaussian method to the results obtained with the Bayesian blocks algorithm used by \citet{nowak12}.
Using this last method, they detected 45 flares, 34 of which where also detected using the Gaussian fitting. This led to a flaring rate of $1.3\pm0.2\,$flares per day.

\citet{yuan16} extended the analysis of \citet{neilsen13} by including the {\it Chandra} observations performed before 2012.
They also improved the detection method by fitting a Gaussian directly on the individual photon arrival times.
This led to the detection of 82 flares, about one-third of which was not detected by the previous method.
The authors deduced a constant observed flaring rate of $1.6\pm0.2\,$flares per day.

\citet{ponti15} studied an even more complete set of X-ray observations by analyzing the {\it Chandra} and {\it XMM-Newton} observations of the Galactic center performed until the end of 2014 as well as the {\it Swift} observations performed in 2014.
They detected for the first time an increase by a factor of 9.3 in the bright flaring rate (i.e., the rate of flares whose the energy is higher than $5\,10^{-9}\,\mathrm{erg\,s^{-1}\,cm^{-2}}$) that occurred from  31 August 2014.

The study of the largest sample of X-ray observations from \sgra{} was conducted by \citet{mossoux17b}, hereafter M\&G17. 
These authors studied the overall observations performed with \textit{XMM-Newton}, \textit{Chandra,} and \textit{Swift} from 1999 to 2015, totalling 9.3\,Ms of exposure and leading to the detection of 107 flares.
After correcting for the observational bias of each instrument, the authors applied the Bayesian blocks algorithm on the continuous flare arrival times to study the temporal distribution of the X-ray flares.
They determined that the overall intrinsic flaring rate is constant from 1999 to 2015, with a value of $3.0\pm0.3\,$flares per day.
By searching for a limit on the flux and energy of the flares that might lead to a change in flaring rate, the authors determined that the flaring rate for the less luminous and less energetic flares decreases between 2013 June and November, and that the flaring rate for the more luminous and more energetic flares begins to increase at the end of 2014 August\ 31.

The most recent study of the flare energy distribution was performed by \citet{bouffard19}.
These authors have analyzed the {\it Chandra} observations from 2012 to 2018 and detected 58 flares. 
They have compared the energy distribution of the flares before and after the potential date at which we may observe the effects of the G2 pericenter passage on the X-ray flares of \sgra{}.
They tested four potential dates in 2014: April 4, May 20, July 16, and August\ 30, that is to say, very close to the pericenter passage of G2 in 2014 April--May \citep{valencias15}.
For each date, they studied the fluence (or energy) distribution of the flares using Monte Carlo simulations.
They showed that the fluence distribution after each date lies within the 70\% confidence level of the simulations and concluded that the characteristics of the flares from \sgra{} did not change.
Their conclusion is consistent with the results of M\&G17 considering all flares.

We here pursue the study made by M\&G17 and analyze additional data from 2016 to 2018 to determine whether the flaring rate has now returned to the level observed before 2013 or if the change in flaring rate has persisted until the end of 2018.
In Sect.~\ref{obs} we report the data reduction from the three telescopes we used in this study, that is to say, \textit{XMM-Newton}, \textit{Chandra,} and \textit{Swift}, before we describe the flare detection methods (Sect.~\ref{flare}).
We then revise the flux--duration distribution of the flares (Sect.~\ref{intrins}).
This allows us to determine updated values of the detection probability for each observation since 1999.
We then compute the intrinsic flaring rate from \sgra{} from 1999 to 2018 taking the sensitivity bias and the errors on flux and duration into account (Sect.~\ref{corr}).
We then search for a flux or fluence range that might lead to a change in flaring rate (Sect.~\ref{rate}).
Finally, we interpret the results in the context of the recent reports on the increase in variability of \sgra{} in the NIR, and we provide preliminary results on the X-ray activity in 2019 (Sect.~\ref{discuss}).

\section{Observations}
\label{obs}
We have analyzed the X-ray observations performed from 2016 February\ 6 to 2018 November\ 2.
The off-axis angle of \sgra{} during the \textit{XMM-Newton} and \textit{Chandra} observations is less than $8\arcmin$, which respects the criterion imposed by M\&G17 to reduce the confusion with the X-ray diffuse emission surrounding \sgra{}.
We kept all the \textit{Swift} observations because the acceptance of an observation is controlled by a correction factor that is computed during the data reduction, which takes the bias induced by the off-axis angle into account.

\subsection{{\it XMM-Newton} observation}
\label{xmm_obs} 

Between 2016 and 2018, \textit{XMM-Newton} \citep{jansen01} has observed the Galactic center on 2016 February\ 26 for an effective exposure of 29.9\,ks (Table~\ref{table:xmm}).
The goal of this observation was to study the short temporal variations in the light curve of the accreting binary SWIFT~J174540.7-290015.
The observation was thus performed in timing mode, which means that \sgra{} was observed only with EPIC/pn \citep{strueder01b}.
The observation was performed with the medium filter.

For the data reduction, we followed the method reported in \citet{mossoux15}.
We used the \texttt{epchain} task of the Science Analysis Software (SAS) package (version 17.0; current calibration files of 2018 September) to extract the event list.
We created the good time intervals (GTI) file, which corresponds to the overall exposure because there is no time range when the soft-proton flare count-rate on the full detector in the 2-10\,keV energy range exceeds $0.009\,\mathrm{count\,s^{-1}\,arcmin^{-2}}$.
We then rejected the dead columns and bad pixels and kept only the single and double events (\texttt{PATTERN $\leq$ 4}).
Finally, we extracted the events from the source+background (src+bkg) region (a disk with $10\arcsec$ radius centered on the radio position of \sgra{}: RA(J2000) = $17^{\mathrm{h}}45^{\mathrm{m}}40\fs036$, Dec(J2000)=$-29^{\circ} 00' 28\farcs17$; \citealt{petrov11}).
Because the absolute astrometry of the EPIC cameras ($1.2\arcsec$; \citealt{EPIC_calibration_status_document}) is smaller than the size of this region, we do not need to register the EPIC coordinates.
The bkg region is a square of about $30\arcsec\times30\arcsec$ whose center is located at about $4\arcmin$ from \sgra{}.
We filtered out the X-ray sources detected with the SAS task \texttt{edetect\_chain} from this region.

\subsection{{\it Chandra} observations}
\label{chandra_obs} 
The Galactic center was observed 16 times with {\it Chandra}/ACIS-S \citep{garmire03} between 2016 and 2018 for a total effective exposure of about 638.1\,ks.
The observing time of \sgra{} with Chandra has thus increased by more than 10\% compared to before 2016.
The log of the {\it Chandra} observations is reported in Table~\ref{table:chandra1}.

We reduced the data using the package called Chandra interactive analysis of observations (CIAO; version 4.9) and the calibration database (CALDB; version 4.7.4).
We first reprocessed the level~1 data with the CIAO script \texttt{chandra\_repro} to create the bad pixel file, flag afterglow events, and filter
the event patterns, afterglow, and bad pixel events.
The src+bkg events were extracted from a disk with $1\farcs25$ radius centered on \sgra{} , while the bkg region is a disk with $8\farcs2$ radius whose center is at $0\farcm54$ south from \sgra{}.

\subsection{{\it Swift} observations}
\label{swift_obs} 
We only selected the {\it Swift} X-ray telescope \citep[XRT;][]{gherels04} observations of the Galactic center that were performed in photon-counting mode.
Between 2016 and 2018, this corresponds to 793 observations of \sgra{} and a total effective exposure of about 690\,ks (PI: N. Degenaar).
This increases the observing time by 45\% compared to before 2016.
The log of the {\it Swift}/XRT observations is reported in Table~\ref{table:swift}.

We followed the improved data reduction method described by M\&G17.
We first applied the \texttt{XRTPIPELINE} task from the HEASOFT software (version 6.24) to filter out the hot and bad pixels and selected grades lower than 12.
We extracted the events located in a disk with $10\arcsec$ radius centered on \sgra{} from these resulting level~2 event lists using the \texttt{XSelect} task.
We did no construct any background region because the orbit of {\it Swift} is below the radiation belts of the Earth, leading to a negligible contamination by the soft-proton flares.
To correct the deformation of the point spread function (PSF) and the vignetting at 2.77\,keV due to the variation in the off-axis angle of \sgra{} between the observations, we used the \texttt{XRTLCCORR} task.
This task computes the correction factors that have to be applied to the light-curve count-rates for each 10\,s interval.
The correction factors also take the position of the source relative to the position of the bad column or bad pixel into account.
When \sgra{} is located on a bad pixel, the correction factors during this observation are high, which unfortunately makes these data useless.
In this study, we considered only observations with a correction factor lower than 3.

\section{Flare detection}
\label{flare}
On 2016 February\ 6, {\it Swift} detected a new X-ray transient named \transquinze{} \citep{reynolds16,degenaar16,corrales17}.
The subsequent {\it Chandra} observation allowed \citet{baganoff16} to determine its position as RA(J2000) = $17^{\mathrm{h}}45^{\mathrm{m}}40\fs664\pm0.3433$, Dec(J2000)=$-29^{\circ} 00' 15\farcs61\pm0.3263$, that is to say, 16$\arcsec$ north from \sgra{}.
Its 2--10\,keV flux is very high, with a long-term variation from 0.5 to $2.5\times 10^{-8}\,\mathrm{erg\,s^{-1}\,cm^{-2}}$.
\citet{ponti16} identified it as a new accreting binary located close to or beyond the Galactic center.
In 2016 May--June, \citet{degenaar16} detected a new X-ray source located even closer to \sgra{} ($10\arcsec$ south), with a 2--10\,keV flux of $(7\pm2)\times10^{-11}\,\mathrm{erg\,s^{-1}\,cm^{-2}}$.
Because these two very active sources are very close to \sgra{}, the apparent flux extracted from the SMBH increases.
Because the efficiency of the detection method strongly decreases as the mean flux increases, the contamination by the transients leads to an increase in the number of undetected flares.
The flare detection method thus needs to be adapted for the observations that were performed during the active phase of the transients.

\subsection{Flares observed with {\it XMM-Newton} and {\it Chandra}}
We used the two-step Bayesian blocks algorithm \citep{mossoux15,mossoux15c} to systematically detect the X-ray flares from the {\it XMM-Newton} and {\it Chandra} observations.
This algorithm is an improvement of the Bayesian blocks algorithm developed by \citet{scargle98} and refined by \citet{scargle13} to detect a statistically significant change in the rate in a time series.
This is an iterative algorithm that iterates on increasingly smaller time ranges until the count rate computed in the considered range is statistically consistent with a constant count rate assuming a Poissonnian distribution of the events.
This method has to be calibrated with a prior number of change points ({\it ncp\_prior}) depending on the mean count rate of the observation and the false-positive rate used.
We assumed a false-positive rate for the flare detection (i.e., two change points) of 0.1\%.
The calibration method is the same as described in \citet{mossoux15}: we constructed sets of 100 event lists with a quiescent (nonflaring) count rate corresponding to those of the studied observation and containing a number of uniformly distributed events that varied with a Poisson statistic.
For each set, we varied the {\it ncp\_prior} from 3 to 9 and recorded the number of false change points that were detected.
The {\it ncp\_prior} was chosen to retrieve the desired false-positive rate.

After the calibration, we applied the Bayesien block method to the observation following \citet{mossoux15}.
We first associated the time of each event with the center of the corresponding frame.
If a number $N$ of events were recorded during the same frame, we considered that $N$ photons were recorded with the same arrival time.
We then merged the GTIs to obtain a continuous flux of events that was divided into Vorono\"i cells.
The start and end time of a cell were defined as the half time between two consecutive photons.
The start of the first cell was defined as the start time of the first GTI, while the stop of the last cell corresponded to the stop time of the last GTI.
We then computed the count rate associated with each Vorono\"i cell as the number of events in the cell divided by its duration.
The count rate was corrected for the integration time of the CCD by applying on each cell a weight defined as the percentage of the recording time compared to the total integration time.

We finally applied the Bayesian blocks algorithm in two steps to correct the observed count-rates for the instrumental background: we first applied the algorithm on the event lists extracted for the src+bkg and bkg regions. 
We then computed a weight as a function of the ratio between the count rates of the blocks detected in the src+bkg and the bkg event lists.
We then applied the algorithm a second time on the src+bkg event list to which we applied the corresponding weight on each Vorono\"i cell.
The resulting optimal segmentation is composed of a succession of blocks describing the count rate of the nonflaring level and the time ranges of the possible flaring activity.
The nonflaring level is defined as the count rate of the longest block, while the count rate of a flare is the mean count rate of the flaring blocks subtracted from the nonflaring level.

At the date of the {\it XMM-Newton} observation, the low-mass X-ray binary \transquinze{} was active.
Owing to the very high luminosity of this active source, a part of its emission was included in the $10\arcsec$ extraction region used for the {\it XMM-Newton} data reduction. 
This leads to an artificial increase by a factor of 30 of the nonflaring level of \sgra{} and thus to a decay in the detection sensitivity of the Bayesian blocks algorithm.
No X-ray flare was detected.

The better angular resolution of the {\it Chandra} telescope means that the contamination from the active transients observed in 2016 was negligible, and the Bayesian blocks algorithm detected nine flares whose characteristics are reported in Table~\ref{table:chandra1}.
One of them ends after the end of the observation.
We converted the mean count rate of these flares into an unabsorbed mean flux based on the conversion factor of $148.2\times 10^{-12}\,\mathrm{erg\,s^{-1}\,cm^{-2}/count\,s^{-1}}$ computed by M\&G17 for an absorbed power law model with the hydrogen column density fixed to $N_\mathrm{H}=14.3\times 10^{22}\ \mathrm{cm^{-2}}$ and a spectral index fixed to $\Gamma=2$.

\subsection{Flares observed with {\it Swift}}
\label{flare_swift}
\begin{figure}[t]
\centering
\includegraphics[trim=5.cm 0.cm 0cm 0cm,clip,width=10cm]{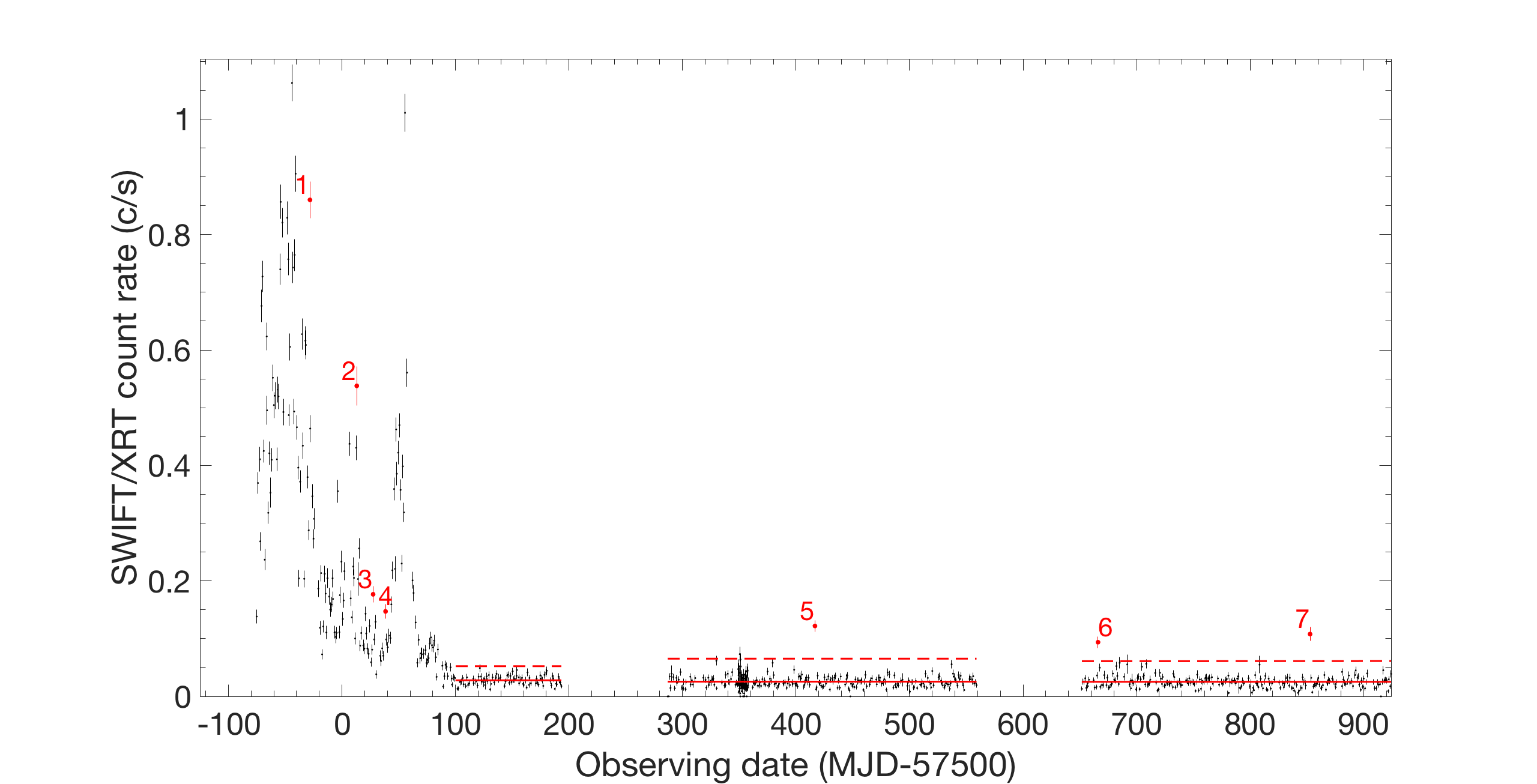}
\caption[2016-2018 \sgra{} light-curve as observed by {\it Swift}.]{Light curve in 2016-2018 of \sgra{} as observed by {\it Swift} (not bkg subtracted). 
The labeled red points are the flares detected by comparison with the two transient light curves before MJD~57600 and by the method proposed by \citet{degenaar13} after MJD~57600.}
\label{swift_lc}
\end{figure}

As mentioned in M\&G17, the method proposed by \citet{degenaar13} is more efficient than the Bayesian block algorithm to detect the flares observed during the short-exposure observations performed by {\it Swift}.
This method is based on the study of a yearly campaign.
We first computed the mean count rate of each observation during a specific year.
Each observation was thus considered as a bin in the yearly light curve.
We then computed the mean count rate and the standard deviation of the yearly campaign.
The mean count rate of the campaign is considered as the nonflaring level in the following.
An observation is finally considered as a flare if the lower limit on its count rate is higher than the yearly mean count rate plus three times the standard deviation.
This method was applied on the 2017 and 2018 campaigns where no active transient was detected close to \sgra{}.
This leads to a detection of three flares (numbers 5 to 7 in Fig.~\ref{swift_lc}).
The net mean count rate of these flares was computed by subtracting the nonflaring level from the mean count rate of the observation.
We then converted the net mean count rate into an unabsorbed mean flux based on the conversion factor of $293.5\times 10^{-12}\,\mathrm{erg\,s^{-1}\,cm^{-2}/count\,s^{-1}}$ computed by M\&G17 for an absorbed power law model with the hydrogen column density fixed to $N_\mathrm{H}=14.3\times 10^{22}\ \mathrm{cm^{-2}}$ and a spectral index fixed to $\Gamma=2$.

During the first half of the 2016 campaign, two X-ray transients entered their active phase.
This leads to a very chaotic light curve, as shown in Fig~\ref{swift_lc}.
It was therefore impossible to use the method proposed by \citet{degenaar13} before 2016 August 1 (MJD~57600).
The alternative method is to compare the light curve of \sgra{}  to an average light curve extracted at the same distance from the transients.
For each transient, we first created ten circular regions with radii of $10\arcsec$ whose centers were at the same distance from the transient as \sgra{}.
\begin{figure}[t]
\centering
\includegraphics[trim=0.cm 0.cm 0cm 0cm,clip,width=8.5cm]{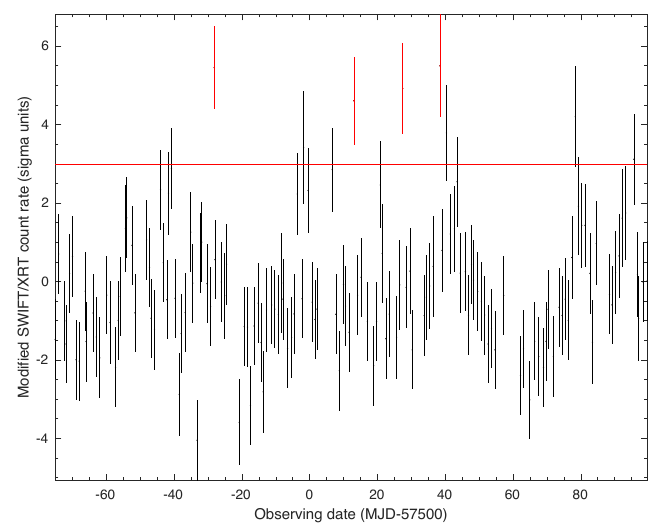}
\caption[Significance of the \sgra{} light-curve compared to those of the transient.]{Significance of the \sgra{} light curve compared to the sum of the averaged light curves of the PSF tail of the transients.
The horizontal line shows the $3\sigma$ limit for the flare detection.
The red points tag the flares from \sgra{}.}
\label{swift_trans}
\end{figure}
We then extracted the event lists for each of these regions and computed the averaged mean count rate and standard deviation for each observation before MJD~57600.
This led to an estimation of the light curve of the PSF tail of the transients at the distance of \sgra{} for the two transients.
The emission from \sgra{} can therefore be considered as an excess compared to the weighted sum of the two averaged light curves.
We flagged an observation when the count rate from the light curve of \sgra{} exceeds those from the sum of the averaged light curves plus three times the standard deviation (Fig.~\ref{swift_trans}).
Four observations were flagged by this method (numbers 1 to 4).
To confirm that these increases in count rate are produced by \sgra{}, we constructed the light curves of these observations using time bins of 100\,s and 2.5\,s, corresponding to the temporal resolution of the XRT camera (see appendix~\ref{swift_flares}).
While the count rate during a \sgra{} flare evolves smoothly, magnetars could produce high single-frame count rates that when binned on a large time bin produce a shape similar to what is observed during a \sgra{} flare.
The analysis of the four light curves shows that only flares 1 and 2 can be attributed to \sgra{} without any doubt.
We therefore only consider these observations as \sgra{} flares in the following.
Their characteristics are reported in Table~\ref{table:swift}.
The unabsorbed mean fluxes of these flares were computed as before, except that the nonflaring level was defined as the weighted sum of the two averaged light curves of the two transients at these observations.

After MJD~57600, the transients were no longer active. 
We therefore used the method of \citet{degenaar13}, but no flare was detected.

\subsection{Comparison with the flares detected by \citet{bouffard19}}
Compared to the {\it Chandra} flares detected by \citet{bouffard19} after 2015, they detected the flares we labeled flares 4 to 8. 
The duration of these flares is slightly shorter that those we derived, but our mean count rate is consistent with the rates they computed after subtracting the quiescent level.
They did not detect the three first flares of 2016 because they excluded the first two {\it Chandra} observations of 2016: the outburst of the low-mass X-ray binary \transquinze\ occurred at that time{}.
We verified that the spectral parameters of the three flares observed on 2016 February\ 13 are consistent with those observed for the previous \sgra{} flares, indicating that these flares are likely emitted by the SMBH and not by the transients (see Appendix~\ref{spectre}).

The authors detected the flare we labeled  flare 9, but they considered it as two separated flares: one with a short duration and very rapid variation, and the other with a longer duration and a lower mean count rate (see the bottom right panel of Fig.~\ref{fig:light_curves}).
The shape of this flare could be the signature of the gravitational lensing of a hotspot orbiting the SMBH.
The first maximum may be due to the gravitational lensing of the light that is emitted by the hotspot when it is located behind the black hole, while the second maximum may by due to the relativistic beaming effect of light that is emitted when the source is moving toward the observer.
Orbital motion like this has previously been observed near the last stable orbit with the NIR GRAVITY-Very Large Telescope Interferometer beam-combining instrument \citep{gravity18b}.
We thus used a ray-tracing code to construct the light curve of a hotspot model \citep{karas92,schnittman04,broderick05,hamaus09,dexter09}.
We modeled the hotspot as a spherical optically thin structure on a Keplerian orbit around \sgra{}.
The spectrum emitted by the hotspot is assumed to be a power law such as $I_\nu\propto\nu^{\alpha}$.
The emission observed in 2--10\,keV is thus the integration of the emitted flux over the frequency range.
Maps of the observed emission were computed using the open-source ray-tracing code \texttt{GYOTO}\footnote{\href{http://gyoto.obspm.fr}{http://gyoto.obspm.fr}} \citep{vincent11}. 
We defined maps of $300\times300$ pixels over one orbital period. 
To compute the light curve, the integration over solid angles is recovered by summing each of these maps over all pixels.

\begin{figure}[t]
\centering
\includegraphics[trim=0.cm 0.cm 0cm 0cm,clip,width=6cm,angle=90]{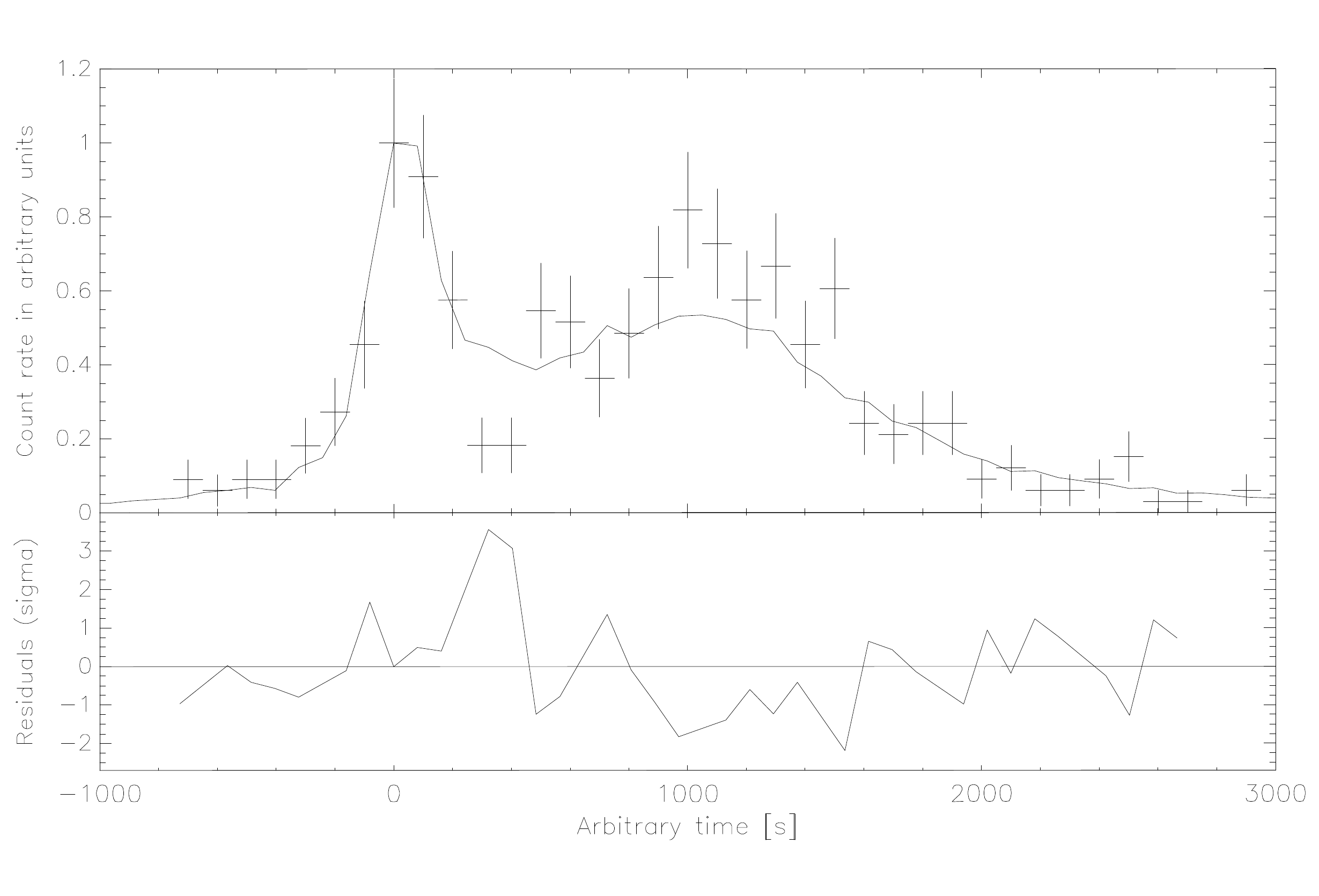}
\caption[The hotspot model.]{Best-fitting hotspot model (solid line) applied to the flare light curve (crosses) observed by \textit{Chandra} on 2018 April 24.
The time bin of the observed light curve is 100\,s.
The bottom panel shows the residuals in units of $\sigma$.}
\label{spot}
\end{figure}

The fitting parameters are the radius of the hotspot, its orbital radius, and the inclination of its orbit.
Because the black hole spin parameter has a very low effect on the light-curve shape, we fixed it to zero.
The time zero of the simulated light curve is defined as the time of the first maximum of the light curve.
The flux of the simulated light curve was also renormalized so that the maximum flux of the simulated curve corresponds to the maximum of the observed light curve.
We fit the observed light curve with this model.
The best-fit curve is shown in Fig.~\ref{spot}\footnote{Because the values of the best-fitting parameters are not representative for the overall characteristics of the accretion flow around \sgra{} and because their physical interpretation is beyond the scope of this paper, we have decided to not mention them to avoid any misinterpretation.}.
We clearly observe that the local minimum between the two maxima of the observed light curve cannot be retrieved, the model remaining at about $3\sigma$ above the observed light curve at this time.
The flux of the second maximum is also difficult to recover, and the model stays at about $2\sigma$ below the observed curve.

The shape of this increase in flux, which is considered as a single flare according to the definition of the Bayesian blocks, is thus difficult to explain as a single flare whose observed flux is shaped by the gravitational lensing of an emitting hotspot orbiting \sgra{}.
To be comprehensive, we therefore also studied the flaring rate considering that this flare is the superposition of two separated flares.

\begin{figure}[t]
\centering
\includegraphics[trim=2.cm 0.cm 0cm 0cm,clip,width=10cm]{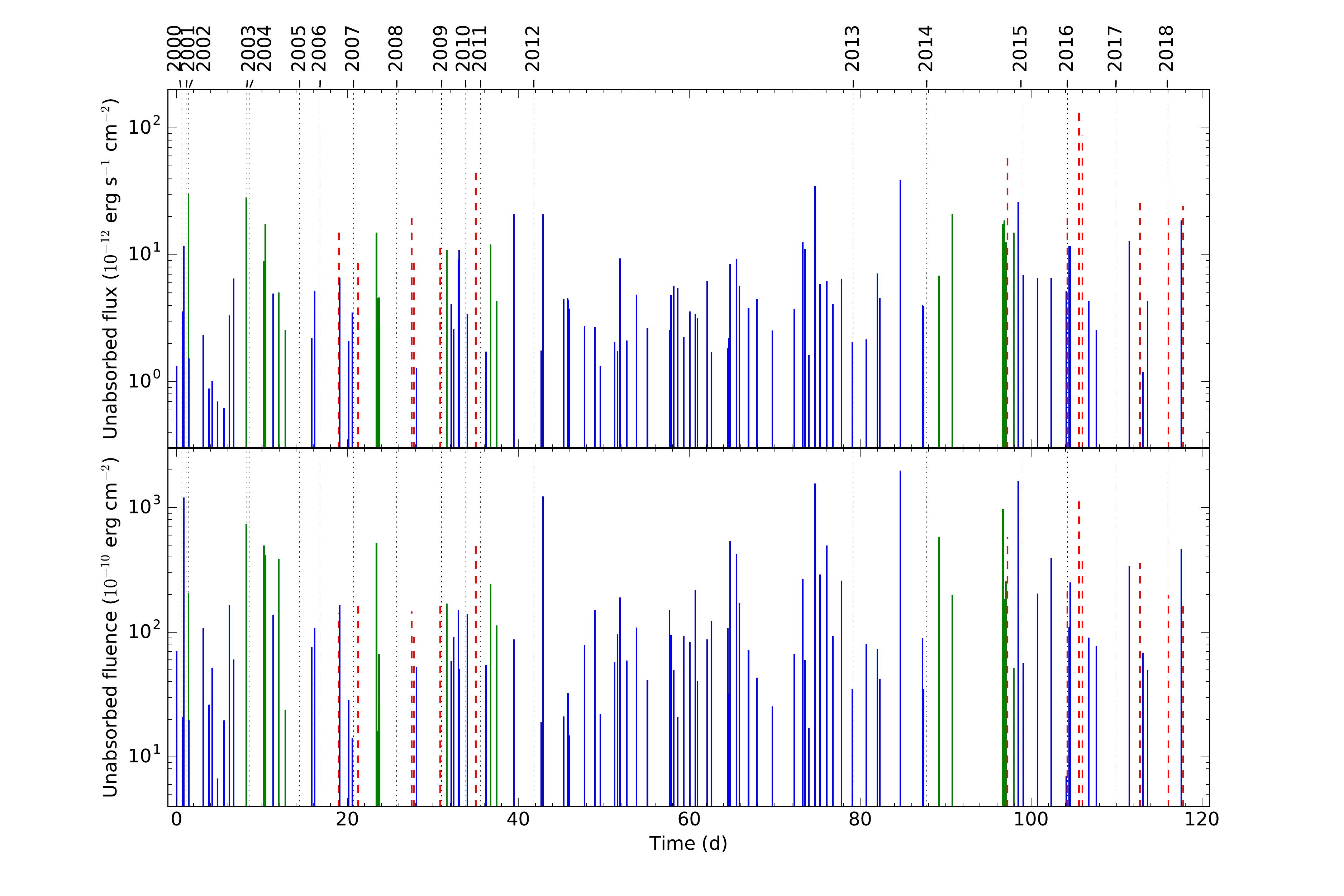}
\caption[Temporal distribution of the flare fluxes and fluences.]{Temporal distribution of the flare fluxes and fluences. 
The mean arrival times of the flares without observing gaps are represented by vertical lines.
The dotted lines are the time of the beginning of the first observation of the year. 
The blue, green, and red lines are the {\it Chandra}, {\it XMM-Newton}, and {\it Swift} flares, respectively. 
The dashed lines tag the flares that begin or end outside the GTIs. 
{\it Top panel:} Mean unabsorbed flux distribution. 
{\it Bottom panel:} Mean unabsorbed fluence (flux time duration) distribution.}
\label{fig:arrival_time}
\end{figure}

\section{Intrinsic flare distribution}
\label{intrins}

\begin{figure*}[t]
\centering
\includegraphics[trim=0.cm 0.7cm 0cm 1.3cm,clip,width=18.cm]{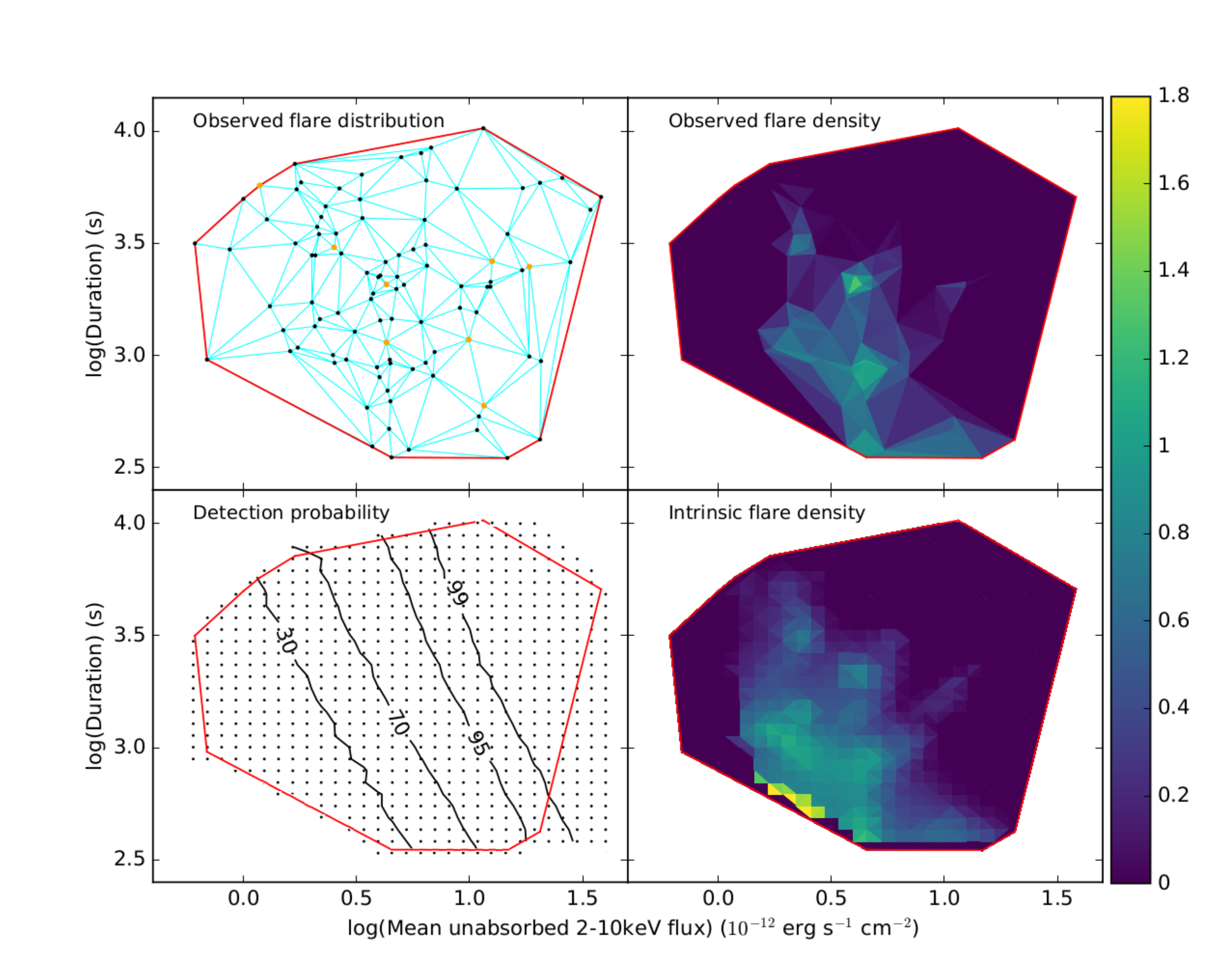}
\caption[Flux-duration distribution of the X-ray flares from \sgra{}]{Flux-duration distribution of the X-ray flares from \sgra{} in a log-log diagram. 
{\it Top left panel:} Flare flux-duration distribution observed with {\it XMM-Newton} and {\it Chandra} from 1999 to 2015 (black dots) and from 2016 to 2018 (orange dots).
The flares beginning before or ending after the GTIs are not represented.
The blue lines are the corresponding Delaunay triangles. 
The red lines define the convex hull. 
{\it Top right panel:} Observed flare density distribution.
The color bar of the filled contours is represented in the right-hand side of the figure in units of $10^{10}\,\mathrm{s^{-1}\,erg^{-1}\,s\,cm^2}$. 
{\it Bottom left panel:} Merged detection efficiency for {\it XMM-Newton} and {\it Chandra} from 1999 to 2018 in percent. 
The dots show the simulation grid. 
{\it Bottom right panel:} Intrinsic flare density distribution corrected for the observing bias computed on the same grid as in the bottom left panel.
The filled contours use the same scale as in the top right panel.}
\label{dens1}
\end{figure*}
We added the 14 flares we detected between 2016 and 2018 to the 107 flares detected before 2016 by M\&G17.
Figure~\ref{fig:arrival_time} shows the 121 flare times without observing gaps over the total exposure time of 119.9$\,$days.
This distribution is highly biased because the flare detection methods are not perfect, which means that only flares are detected whose flux is high enough and that last for long enough.
Moreover, given the same detection method, the detection probability of a given flare is different from one observational instrument to an other because of the difference in the angular resolution and efficiency of the CCD. 
To compare the flaring rates that are observed with different instruments at different times, we therefore corrected for the detection efficiency of each observation.

\subsection{Revised method of M\&G17}

\begin{figure*}[t]
\centering
\includegraphics[trim=0.cm 0.cm 2cm 0cm,clip,width=9.cm]{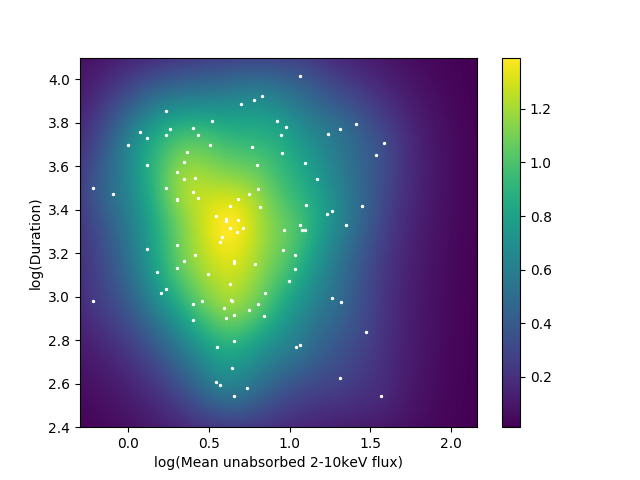}
\includegraphics[trim=0.cm 0.cm 2cm 0cm,clip,width=9.cm]{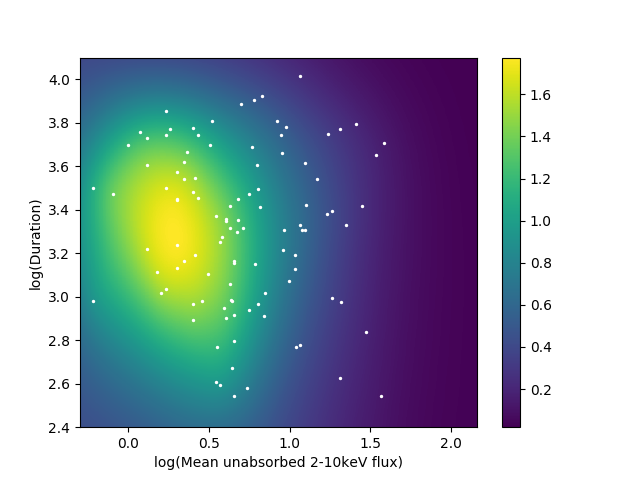}
\caption[Flux-duration distribution of the X-ray flares from \sgra{} using DIVA]{Flux-duration distribution of the X-ray flares from \sgra{} in a log-log diagram using DIVA. 
The color bars represented in the right-hand side of each panel are in units of $10^{10}\,\mathrm{s^{-1}\,erg^{-1}\,s\,cm^2}$. 
{\it Left panel:} Flare flux-duration distribution observed with {\it XMM-Newton} and {\it Chandra} from 1999 to 2018.
The flares beginning before or ending after the GTIs are not represented.
{\it Right panel:} Intrinsic flare density distribution corrected for the observing bias and the error on the flux and duration.
}
\label{dens2}
\end{figure*}
As a first analysis, we used the method developed by M\&G17 to determine the intrinsic flare distribution.
The main steps are reported here.
We first add the 8 {\it Chandra} flares (the flare that ended after the end of the observation was discarded) to the 99 flares that were previously detected as a whole with {\it Chandra} and {\it XMM-Newton} in the diagram representing the flares as a function of their duration and flux.
Most of the flares are located in the bottom left part of the diagram, meaning that most flares have short durations and low fluxes (see Fig.~5 of M\&G17).
As an update to the M\&G17 method, we therefore decided to work in the log-log diagram.
We then constructed the Delaunay triangles (top left panel of Fig.~\ref{dens1}) and used the Delaunay field estimator (DTFE; \citealt{schaap00,weygaert09}) to compute the observed density per surface unit associated with each flare.
We note that this density $d_\mathrm{log}$ is computed in the log-log diagram.
The physical density $d$ per unit of flux and duration at a flux $F$ and a duration $T$ is thus retrieved by
\begin{equation}
   d_\mathrm{obs}=\frac{d_\mathrm{log}}{10^{(F+T)}\,(\ln{(10)})^2}\, .
\end{equation}
The density $d_\mathrm{log}$ was then linearly interpolated inside the convex hull to determine the observed flare density (top right panel of Fig.~\ref{dens1}).

We then computed the local detection efficiency ($p_\mathrm{nf}(\pmb{x})$) for each {\it XMM-Newton} and  {\it Chandra} observation from 2016 to 2018.
To do this, we used the same simulation grid as defined in M\&G17 (points in the bottom left panel of Fig.~\ref{dens1}) and created for each point $\pmb{x}$ of the grid 100 simulated event lists whose nonflaring level corresponds to those measured for each observation. 
We added a Gaussian flare to each event list whose duration (defined as twice the Gaussian standard deviation) and flux correspond to those of the grid points, as explained in Appendix~D of M\&G17.
We then applied the Bayesian block algorithm to determine how many times the flare was detected.
The merged local detection efficiency ($p(\pmb{x})$; contours in the bottom left panel of Fig.~\ref{dens1}) was constructed by averaging the resulting grids of the local detection efficiency $p_\mathrm{nf}(\pmb{x})$ with those previously computed for the observations before 2016 by M\&G17.

The intrinsic flare distribution (bottom right panel of Fig.~\ref{dens1}) was finally retrieved by correcting the observed flare distribution for the merged local detection efficiency with $d_\mathrm{intr}(\pmb{x})=d_\mathrm{obs}(\pmb{x})/p(\pmb{x})$ (Eq.~17 of \citealt{weygaert09}).

\subsection{Revised method of determining the intrinsic flare distribution}
As a second attempt, we used the data-interpolating variational analysis (DIVA) software developed by the GHER group of the University of Li\`ege \citep{brasseur94,brasseur96,troupin12}.
This software aims to reconstruct a continuous field from irregularly spaced data that might be affected by errors.
We used the n-dimensional tool (DIVAnd, version 1.0), which interpolates the data on a curvilinear grid in any number of dimensions \citep{barth14}.
In this study, we work on two dimensions: flux and duration.

We again worked on a log-log diagram.
One of the most important parameters of the DIVA software is the correlation length $L,$ which is an estimate of the distance over which a data point influences its neighborhood; in other words, over which distance a point is representative. 
In the present context an increase in this length scale spreads the calculated density over a larger distance but decreases its maximum value accordingly.
To estimate the observed density, the value of $L$, computed for the flux and duration axes using the empirical rule of thumb of \citet{silverman86}, is the average distance to the mean value of the flux and duration, respectively, divided by the number of flares to the exponent 1/6.
It is thus a constant value of about 0.18 along the log-flux axis and 0.17 along the log-duration axis.
The observed density ($d_\mathrm{obs}$; left panel of Fig.~\ref{dens2}) is the sum of the density fields interpolated on a grid of 500x500 points considering each flare individually as a unity point.

Using this density, we were able to compute a scaling of the correlation length for each point $\pmb{x}$ of the grid as a function of the observed density. This technique is called adaptive or variable kernel density estimation:
\begin{equation}
   S(\pmb{x})=\frac{\int\!\int (d_\mathrm{obs}(\pmb{x}))^{0.5}\pmb{x}}{(d_\mathrm{obs}(\pmb{x}))^{0.5} \, \int\!\int \pmb{x}}\, .
\end{equation}
The mean value of $S(\pmb{x})$ is around one, with higher values where the density is lower.

We then took the error on the duration and flux of the flares into account.
We computed the error on the flux and duration of each observed flare as follows:
we reproduced each flare by simulating event lists with a Poisson flux reproducing the corresponding nonflaring level plus a Gaussian-flare light curve with the same mean count rate and duration as observed.
We then applied the Bayesian blocks algorithm and recorded the resulting duration and mean count rate (we simulated event lists to reach 100 detections of the considered flare).
The error on the duration ($T_\mathrm{err}$) and mean count rate (later converted into flux $F_\mathrm{err}$) are defined as the standard deviation of these values.
The errors of the overall flares were then interpolated on the grid using DIVA.
These errors affect the flare density by changing the correlation length.
The larger the errors, the more spread an observation by increasing the local length scale accordingly.
The value of the correlation length for each point $\pmb{x}$ of the interpolation grid is therefore
\begin{equation}
  \begin{aligned}
   L_F(\pmb{x})&= 0.18\,S(\pmb{x})+F_\mathrm{err}\\
   L_T(\pmb{x})&= 0.17\,S(\pmb{x})+T_\mathrm{err}\, .
  \end{aligned}
\end{equation}

Finally, the intrinsic flare density was computed by considering each flare associated with its detection probability interpolated at the flare position from the merged local detection efficiency presented in the bottom left panel of Fig.~\ref{dens1}.
Each flare was therefore no longer considered as a unity point, but as the inverse of its detection probability (computed between 0 and 1).
The resulting intrinsic flare density $d_\mathrm{intr}$ is shown in the right panel of Fig.~\ref{dens2}.

\begin{figure}[t]
\centering
\includegraphics[trim=0.cm 0.cm 0cm 0cm,clip,width=8.5cm]{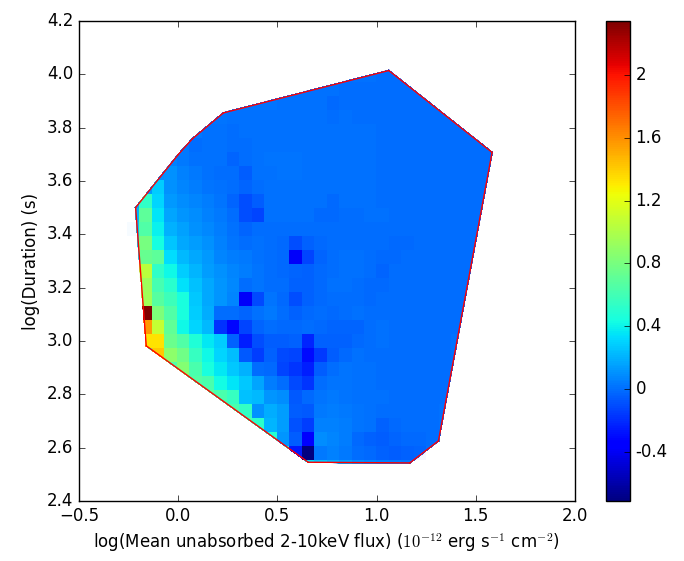}
\caption[Difference between the two methods.]{Difference in the intrinsic density between the methods of M\&G17 and DIVA.
The color bar is in units of $10^{9}\,\mathrm{s^{-1}\,erg^{-1}\,s\,cm^2}$.}
\label{diff}
\end{figure}
Compared to the M\&G17 method, the DIVA method determines an intrinsic density that is higher for the very low luminosity flares (Fig.~\ref{diff}).
However, the difference in density is small (less than $2.5\times10^{9}\,\mathrm{s^{-1}\,erg^{-1}\,s\,cm^2}$) and only affects the flares that are detected with a low probability.
The weight of these points is therefore low in the rest of this study.

\section{Correction of the observed temporal distribution of the flares}
\label{corr}
To correct the observed flaring rate for the detection bias, we used the intrinsic flare distribution ($d_\mathrm{intr}(\pmb{x})$) computed in the previous sections and the local detection efficiency ($p_\mathrm{nf}(\pmb{x})<1$) of each observation.
We computed an average flare detection efficiency $\eta<1$ inside the convex hull, which was used to weight the exposure of each observation as done in M\&G17,
\begin{equation}
   \eta=\frac{\int\!\int d_\mathrm{intr}(\pmb{x})\,p_\mathrm{nf}(\pmb{x})\,d\pmb{x}}{\int\!\int d_\mathrm{intr}(\pmb{x})\,d\pmb{x}}\, .
   \label{eq:eta}
\end{equation}
For the {\it Swift} observations after MJD~57600, the local detection efficiency was computed using the \citet{degenaar13} detection method instead of the Bayesian block algorithm.
The flares that occurred before MJD~57600 were detected using a $3\sigma$ limit in the comparison of the \sgra{} light curve and the weighted sum of light curves of the PSF tail of the two transients.
For each observation, we therefore have a strict limit on the flux that the flare might have to be detected at more than $3\sigma$.
The local flare detection efficiency of these observations is thus a bimodal distribution (always detected or never detected) that only depends on the flux of the flare.

When we compare the M\&G17 and DIVA methods for determining the intrinsic flare distribution, the value of $\eta$ is different by less than 2\% for each observation.
Because the DIVA method is more accurate because it includes the error on the flux and duration in the calculations, we computed the average detection efficiency using this last method in the rest of this paper. 

\begin{figure}[t]
\centering
\includegraphics[trim=2.cm 0.cm 0cm 0cm,clip,width=10.cm]{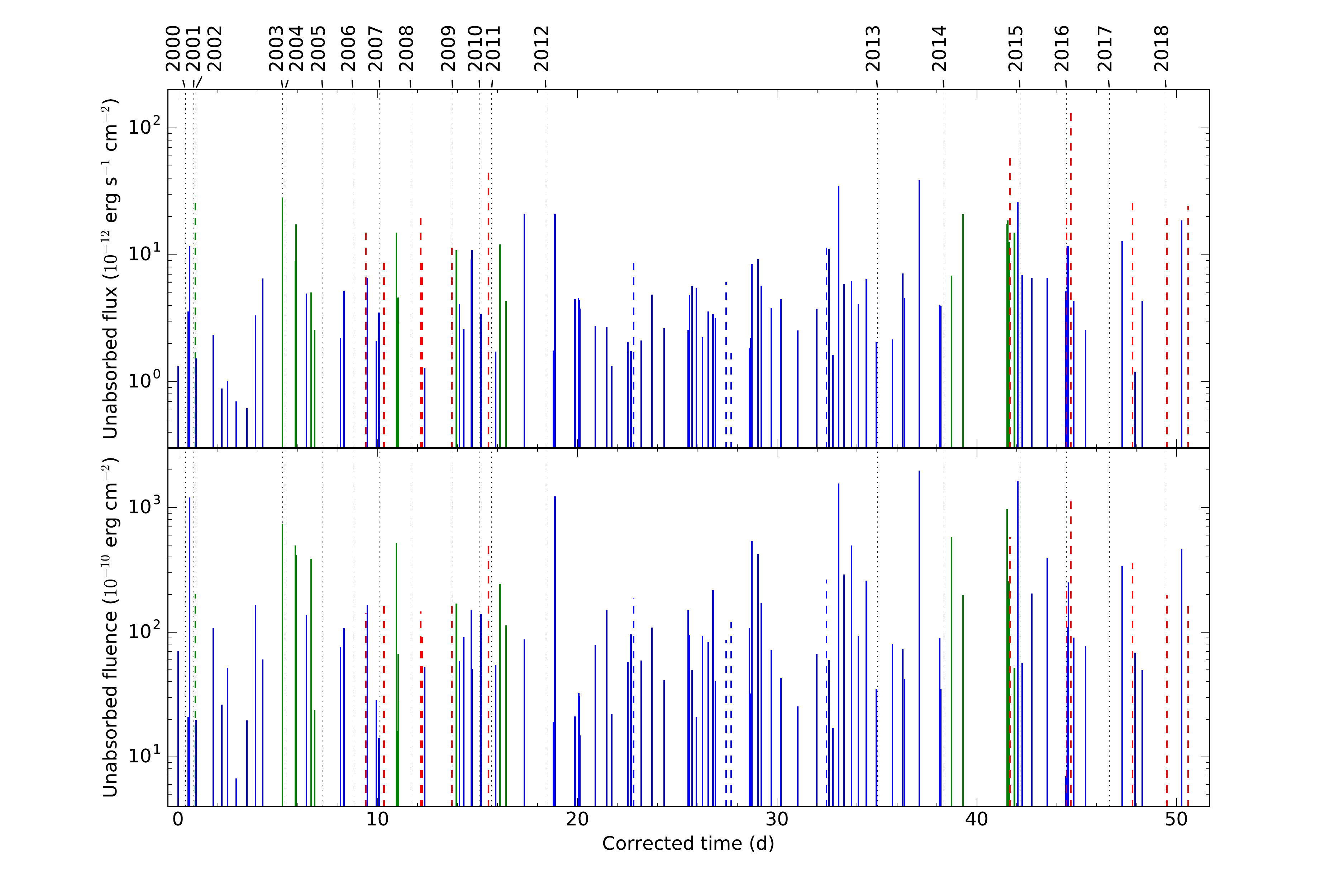}
\caption[Corrected temporal distribution of the flare fluxes and fluences.]{Temporal distribution of the flare fluxes and fluences corrected for the sensitivity bias.
See Fig.~\ref{fig:arrival_time} for details.}
\label{fig:arrival_time_corr}
\end{figure}

For each observational exposure $T$, we computed the corrected observational exposure as $T_\mathrm{corr}=T\,\eta$ . This led to a higher and unbiased flaring count rate in the corresponding observation.
Figure~\ref{fig:arrival_time_corr} shows the 121 flare times without observing gaps over the total corrected exposure time of 51.15$\,$days.

\section{Study of the unbiased flaring rate}
\label{rate}
As a first study, we considered the 121 flares that are distributed over the corrected exposure of 51.15$\,$days.
The search for a change of flaring rate was performed using the Bayesian blocks method.
We first divided the corrected exposure into Vorono\"i cells defined to contain only one flare each.
We then applied the Bayesian blocks algorithm with a false-positive rate of 0.05, leading to a significance of 95\% of any change in the detected flaring rate.
We calibrated the {\it ncp\_prior} considering a uniformly distributed arrival time of 121 flares during 51.15$\,$days.
The corresponding {\it ncp\_prior} is 7.7.
The algorithm did not detect any change in flaring activity, as was observed by M\&G17.
The intrinsic flaring rate is thus $2.4\pm0.2$ flares per day.
This result is different at $1.7\sigma$ from the $3.0\pm0.3$ flares per day computed by M\&G17.
This difference comes from an update in the intrinsic flare distribution that leads to an increase in the values of $\eta$ and thus to an increase in the corrected exposure time. It is not due to a flaring rate that is averaged with a possibly lower flaring rate after 2016.
The intrinsic flaring rates before 2016 (107 flares) and after 2016 (14 flares) are very similar.

\subsection{Search for a flux threshold}
As in M\&G17, we then searched for a flare flux range that might lead to a change in flaring rate.
An additional improvement of the method used by M\&G17 can also be applied here.
When a flux or fluence (flux times duration) threshold is searched for, M\&G17 used the same value of $\eta$ as computed considering the overall flares.
However, considering only a given range of flux or fluence, the corresponding average flare detection efficiency is different.
For example, when we consider only flares with a flux higher than $35\times10^{-12}\,\mathrm{erg\,s^{-1}\,cm^{-2}}$, the bottom left panel of Fig.~\ref{dens1} shows that almost all these flares are detected with a probability of 100\%.
This means that we miss fewer flares with a flux higher than $35\times10^{-12}\,\mathrm{erg\,s^{-1}\,cm^{-2}}$ than those with a flux lower than $15\times10^{-12}\,\mathrm{erg\,s^{-1}\,cm^{-2}}$ , for example.
The value of the average flare detection efficiency must therefore be computed for each observation, but also for each flux and fluence threshold by applying Eq.~\ref{eq:eta} on the corresponding flux or duration grid.

We first applied the top-to-bottom iterative search, where the arrival time of the flare with the highest flux is removed at each step while we kept the exposure time of the corresponding observation fixed.
At each step, we therefore computed an updated value of $\eta$ corresponding to the flux limit and leading to an updated value of the corrected arrival times.
We then computed the Vorono\"i cells and the calibration of the {\it ncp\_prior}.
We finally applied the Bayesian blocks algorithm and iterated until a change in flaring rate was found.
At each step, the values of $\eta$ decrease because we considered flares with increasingly lower flux, implying increasingly lower local detection efficiencies.
The corrected time therefore decreases at each step with the number of flares.
This leads to a flaring rate that decreases more slowly at each step than reported by M\&G17.
No change in flaring rate was found by the Bayesian blocks algorithm.

M\&G17 found a decay in the flaring rate for flares with a flux lower than $6.5\times10^{-12}\,\mathrm{erg\,s^{-1}\,cm^{-2}}$ between 2013 May 25 and July 27.
After 2016, six flares satisfy this flux limit.
When we use the new method to compute $\eta$, the flaring rates before 2013 May 25, between 2013 July 27 and 2015 November 2, and after 2015 November 2 are $2.9\pm0.4$, $1.4\pm0.6$, and $1.8\pm0.7$ flares per day, respectively.
After 2013 July 27, the two flaring rates are constant within the error bars, but the difference between the flaring rates before and after this date is not significant enough to be considered as a change in flaring rate.
The change in flaring rate found in M\&G17 is therefore probably due to a calculation of $\eta$ that was not adapted to the flux range considered at each step.

\begin{figure}[t]
\centering
\includegraphics[trim=2.8cm 0.cm 0cm 12cm,clip,width=9.5cm]{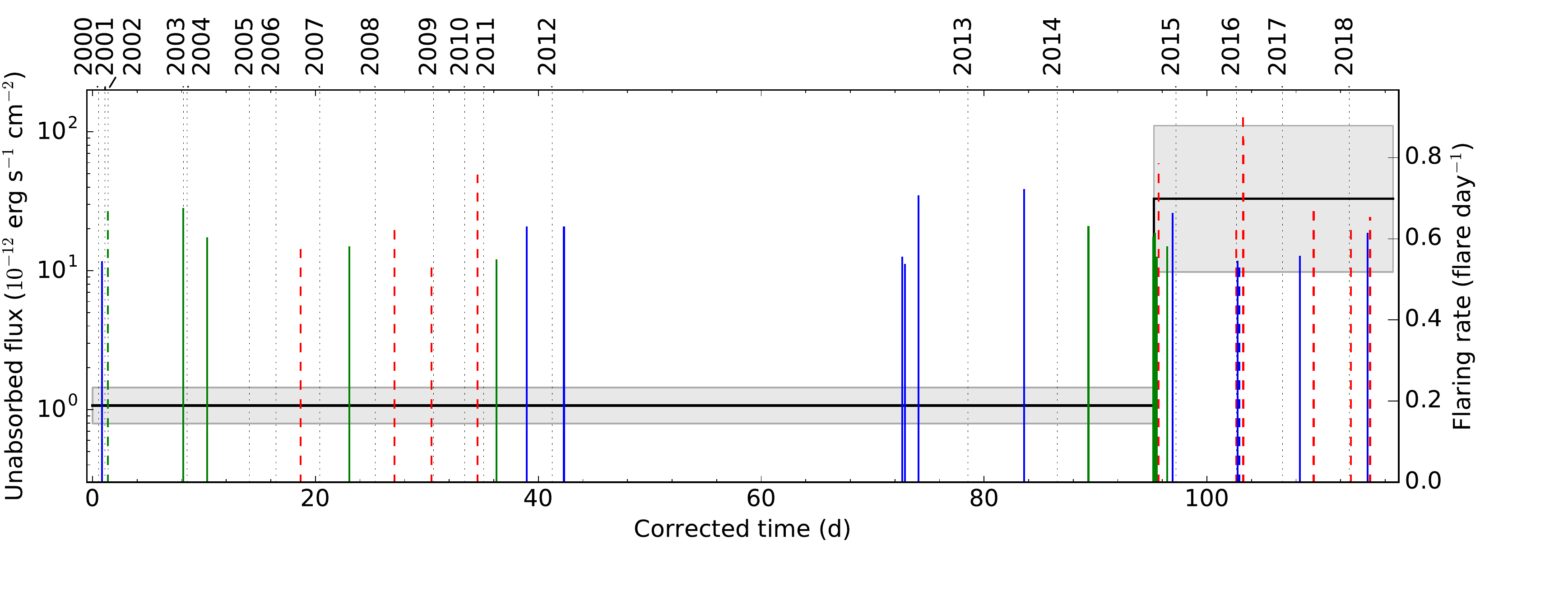}
\caption[Change of flaring rate while studying the flux.]{X-ray flaring rate from 1999 to 2018 computed by the Bayesian blocks algorithm for flares with fluxes higher than $11.0\times10^{-12}\,\mathrm{erg\,s^{-1}\,cm^{-2}}$. 
See Fig.~\ref{fig:arrival_time} for details.
The Bayesian blocks are indicated with thick black lines, and their errors are the horizontal gray rectangles.
}
\label{fig:flux}
\end{figure}

We then applied the bottom-to-top iterative search where the arrival time of the flare with the lowest flux is removed.
At each step, the values of $\eta$ now increase, leading to a longer corrected time, while the number of flares decreases.
The flaring rate therefore decreases faster at each step than reported by M\&G17.
A change in flaring rate was found for flares whose flux was higher than $11\times10^{-12}\,\mathrm{erg\,s^{-1}\,cm^{-2}}$ (see Table~\ref{flaring_rate}).
This flux limit is higher than the limits found by M\&G17, who determined a limit of $4.0\times10^{-12}\,\mathrm{erg\,s^{-1}\,cm^{-2}}$.
The first block contains 18 flares and the last encloses 15 flares (see Fig.~\ref{fig:flux}).
The increase in flaring rate occurs between the two {\it XMM-Newton} flares of 2014 August\ 30.
This date is the same as the date found by M\&G17.
We therefore confirm the increase in flaring rate for the brightest (or most luminous) flares.

\subsection{Search for a fluence threshold}
We then searched for a flare fluence range that  would lead to a change in flaring rate.
As previously, we applied a top-to-bottom and a bottom-to-top iterative search with a calculation of the value of $\eta$ for each observation at each step.

No change in flaring rate was found by the Bayesian blocks algorithm for the top-to-bottom search.
This result is different from the results of M\&G17, who found a decrease in the flaring rate from 2013 July 27--October\ 28 with a significance of 95.1\%.
As for the flux, this discrepancy is likely due to the values of $\eta$ that were not adapted in M\&G17.

\begin{figure}[t]
\centering
\includegraphics[trim=2.8cm 0.cm 0cm 12cm,clip,width=9.5cm]{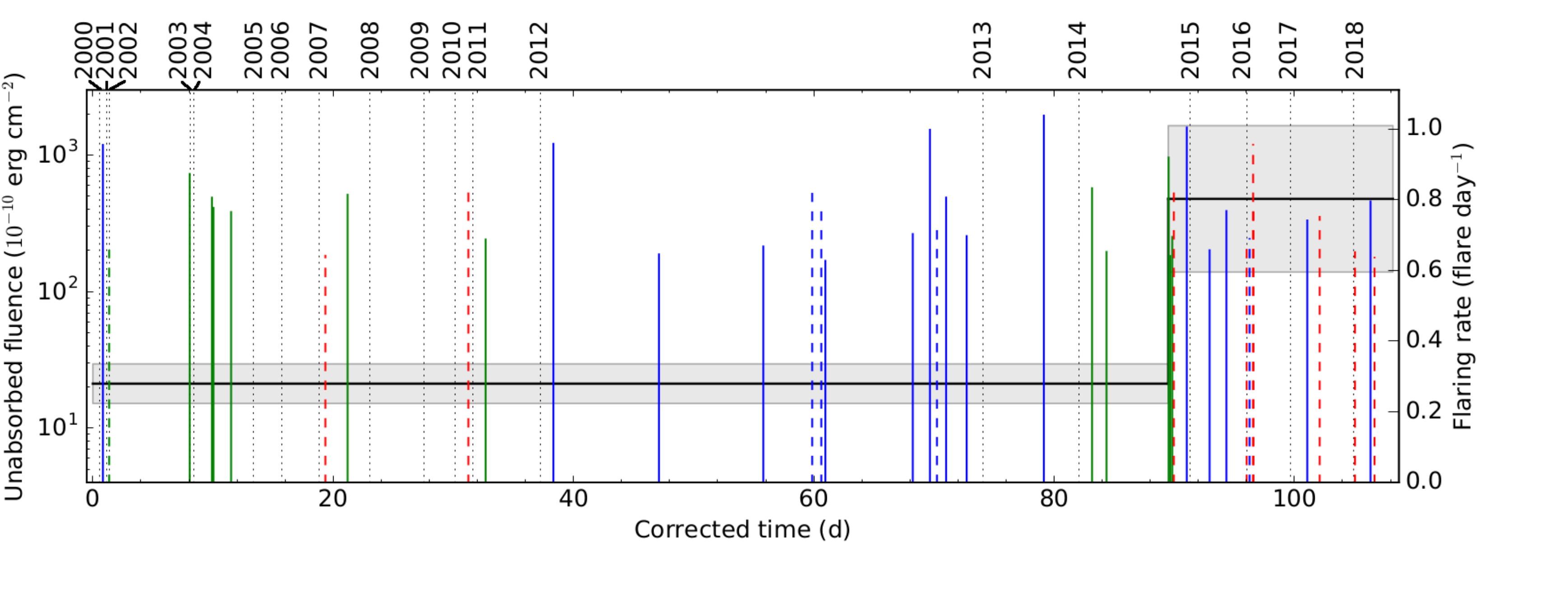}
\caption[Change of flaring rate while studying the fluence.]{X-ray flaring rate from 1999 to 2018 computed by the Bayesian blocks algorithm for flares with fluences higher than $167.9\times10^{-10}\,\mathrm{erg\,cm^{-2}}$. 
See Fig.~\ref{fig:arrival_time} for details.
The Bayesian blocks are indicated with thick black lines, and their errors are horizontal gray rectangles.
}
\label{fig:fluence}
\end{figure}

For the bottom-to-top search, we found that an increase in flaring rate occurred at the same date as in the previous section.
The first block contains 25 flares and the second blocks has 15 flares (see Fig.~\ref{fig:fluence}).
The date of this change in flaring rate is very similar to the dates detected by M\&G17 (one day before), but the limit on the fluence is higher than $91.3\times10^{-10}\,\mathrm{erg\,cm^{-2}}$ found by these authors (see Table~\ref{flaring_rate}). 
We therefore confirm the increase in flaring rate for the most energetic flares.

\begin{table}
\caption{Summary of the change in X-ray flaring rates detected between 1999 and 2018 \label{flaring_rate}}
\begin{center}
\resizebox{0.493\textwidth}{!}{
\begin{tabular}{@{}lcc@{}}
\hline
\hline
  & Top-to-bottom & Bottom-to-top \\
\hline
Flux thresholds ($10^{-12}\,\mathrm{erg\,s^{-1}\,cm^{-2}}$) & No threshold & $\geq 11.0$ \\
Number of less and most luminous flares & 121 & 33 \\
First block (flares per day) & $2.4\pm0.2$ &  $0.19\pm0.04$\\
Corrected time of the change point & \dots\dots\dots & 95.2 \\
Date of the change point & \dots\dots\dots & 2014 Aug.\ 30 \\
Second block (flares per day) & \dots\dots\dots & $0.70\pm0.18$ \\
Significance (\%) & \dots\dots\dots &  97\\
$\mathrm{ncp\_prior}$ & \dots\dots\dots & 6.8 \\
\hline
Fluence thresholds ($10^{-10}\,\mathrm{erg\,cm^{-2}}$) & No threshold & $\geq 167.9$ \\
Number of less and most energetic flares & 121 & 40 \\
First block (flares per day) & $2.4\pm0.2$ &  $0.28\pm0.06$\\
Corrected time of the change point & \dots\dots\dots & 89.6 \\
Date of the change point & \dots\dots\dots & 2014 Aug.\ 30--31 \\
Second block (flares per day) & \dots\dots\dots & $0.8\pm0.2$ \\
Significance (\%) & \dots\dots\dots &  96\\
$\mathrm{ncp\_prior}$ & \dots\dots\dots & 6.3 \\
\hline
\end{tabular}
}
\end{center}
\end{table}

\section{Discussion}
\label{discuss}
Considering the overall flares, our results therefore confirm those found by M\&G17 and by \citet{bouffard19}.
A change in flaring rate was only found considering a range of flux or fluence.
As mentioned in Sect.~\ref{flare}, \citet{bouffard19} considered {\it Chandra} flare 9 as two separated flares.
We therefore also applied our analysis considering this flare as two flares with durations of 791 and 2349~s and mean unabsorbed fluxes of 21.7 and $20.8\times10^{-12}\,\mathrm{erg\,s^{-1}\,cm^{-2}}$.
However, this did not change our global result: a change in flaring rate is found for the brightest and most energetic flares at the same date as was found in the previous section.
The lower limit on the flux is lower than previously ($9.2\times10^{-12}\,\mathrm{erg\,s^{-1}\,cm^{-2}}$), leading to the addition of four flares in the first block and two flares in the second block, but the count rates of the blocks remain within the error bars.
For the fluence, the lower limit is the same.
Once again, the count-rates of the blocks remain consistent within the error bars because the only difference is flare 9, which is now considered as two flares.

The increase in flaring rate of the most energetic fares must be placed in the context of the recent increase in activity observed from \sgra{} in NIR and X-rays.
\citet{do19} observed the Galactic center with the Very Large Telescope (VLT) and the Keck Telescope in 2019 April and May.
They recorded the highest variability ever observed over 20 years of NIR observations (up to a factor of 75 within two hours).
They also detected the largest flare ever observed (more than 6~mJy in the $K'$ band).
The overall distribution of the flux variation during the four observation nights shows a large deviation compared to historical data, especially in the high flux tail of the distribution.
This change in activity in 2019 may first seem difficult to relate with the increase in bright X-ray flaring rate on 2014.
However, this change in  activity in NIR was revealed by comparison with the complementary cumulative distribution function (CCDF) model computed by \citet{witzel18} for the VLT and Keck observations between 2003 and 2013.
There is thus a lake of VLT and Keck observations between 2013 and 2019 that may reveal an increase in the NIR activity at a date close to the dates observed in X-rays.
\citet{witzel18} have also analyzed the {\it Spitzer/IRAC} data performed in 2013, 2014, 2016, and 2017.
However, these authors studied this dataset as a whole to create a CCDF model for this instrument.
In a future work, it may be interesting to divide this dataset into two sets (e.g., 2013--2014 and 2016--2017) to compare the parameters of the CCDF models and to study the change in model parameters.

\begin{figure}[t]
\centering
\includegraphics[trim=0cm 0.cm 0cm 0cm,clip,width=3cm,angle=90]{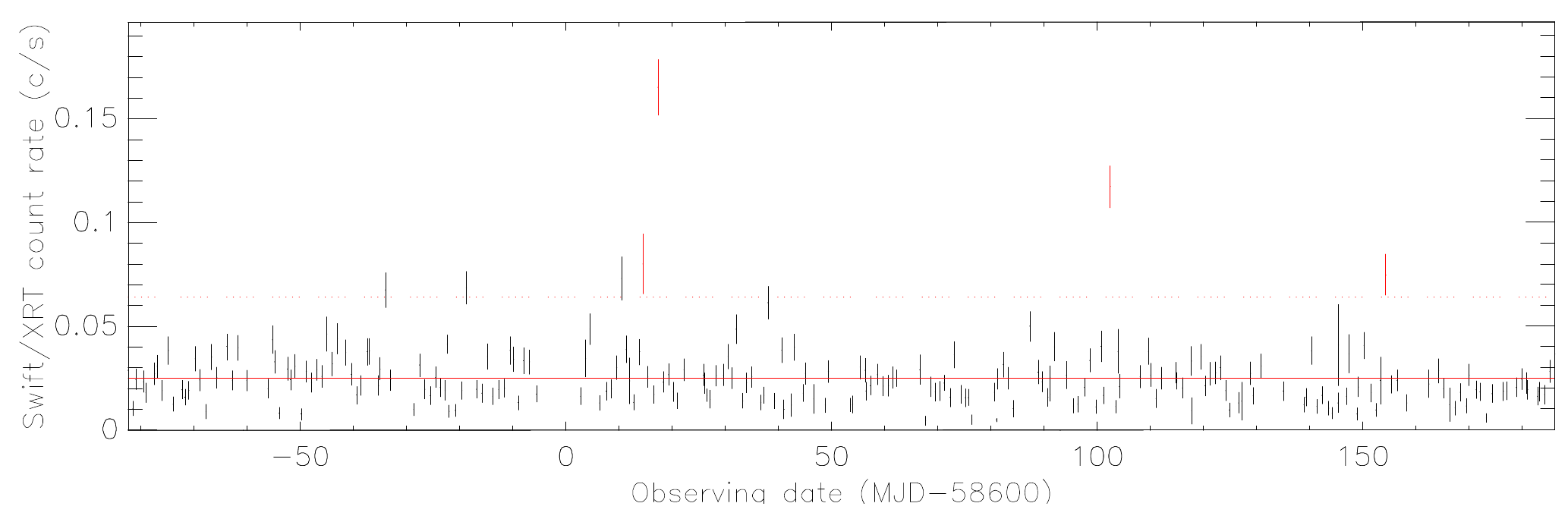}
\caption[2019 \sgra{} light-curve.]{2019 light curve of \sgra{} as observed by {\it Swift} (not bkg subtracted). 
The red points are the flares detected with the method proposed by \citet{degenaar13}.}
\label{camp_19}
\end{figure}

In X-rays, we mention the results obtained with the ART-XC telescope on board the {\it Spektr-RG} orbital observatory, which observed the Galactic center in 5--16\,keV on 2019 August and detected an increase in activity and a variability on a timescale of a few kiloseconds to a few hours \citep{pavlinsky19a,pavlinsky19b}.
Moreover, \citet{degenaar19a} and \citet{reynolds19a,reynolds19b} claimed that in 2019, Swift observed three flares with a flux higher than $10^{-11}\,\mathrm{erg\,cm^{-2}}$ during this campaign.
To verify this result, we analyzed the {\it Swift}/XRT data obtained in 2019 as explained in Sect.~\ref{swift_obs} and \ref{flare_swift}.
We detected four flares (see Fig.~\ref{camp_19}).
The first three flares are the flares detected by \citet{degenaar19a} and \citet{reynolds19a,reynolds19b}.
This is the largest number of flares ever observed by {\it Swift} during a yearly campaign.
The mean number of flares observed by {\it Swift} during a yearly campaign between 2006 and 2018 is $0.93\pm0.26$ flares, and only the 2008 campaign led to the detection of three flares.
We added the four {\it Swift} flares to the list of 121 flares that we detected until 2018.
Based on analyzing the flaring rate, the high flaring rate state is persistent considering the {\it Swift} flares of 2019.
The 31 observations performed with {\it Chandra} and {\it XMM-Newton} in 2019 (which will be public beginning in March 2020) will be very useful for covering a wider range of flare fluxes and durations and thus for drawing more stringent constraints on the persistence of the increase in flaring rate for the brightest and most energetic flares.

\section{Conclusion}
\label{concl}
We analyzed the {\it Swift}, {\it XMM-Newton,} and {\it Chandra} observations of the Galactic center from 2016 to 2018 to pursue the study performed by \citet{mossoux17b} of the X-ray activity of the supermassive black hole \sgra{}.
We detected 14 additional flares (9 with {\it Chandra}/ACIS-S and 5 with {\it Swift}) that we added to the 107 flares that were previously detected.
We developed a new detection method for the flares observed by {\it Swift} at the beginning of 2016 because two very active X-ray transients were present close to \sgra{} at these dates.
Of the 5 flares detected with {\it Swift, 2} were tagged with this method.
We also revised the method used by \citet{mossoux17b} to determine the intrinsic flare distribution and to take the errors on the flare fluxes and durations into account.
We used this intrinsic flare distribution and the flare detection efficiency of each observation to determine an intrinsic distribution of the flare arrival times.

We studied this distribution with the Bayesian blocks algorithm, but no significant change in flaring rate was found for the 121 X-ray flares (intrinsic flaring rate of $2.4\pm0.2$ flares per day).
We then studied the flaring rate for flares lying in a given range of flux and fluence by iteratively applying the Bayesian blocks algorithm after rejecting the less or more luminous or energetic flares at each step.
As an update of the \citet{mossoux17b} method, we computed at each step the average flare detection efficiencies corresponding to the considered flux and fluence range.
We did not detect any significant change in flaring rate for the less luminous and less energetic flares, in contrast to the result of \citet{mossoux17b}.
However, we identified an increase in the flaring rate by a factor of about three for the brightest and most energetic flares (flux higher than $11\times10^{-12}\,\mathrm{erg\,s^{-1}\,cm^{-2}}$ and fluence higher than $1.68\times10^{-8}\,\mathrm{erg\,cm^{-2}}$).
These changes in flaring rate that occurred since the {\it XMM-Newton} flares of 2014 August\ 30 confirm the results of \citet{mossoux17b}.

An increase in the \sgra{} variability has also been detected in the NIR between the VLT and Keck data obtained in 2003--2013 and the Keck data obtained in 2019.
Moreover, several Astronomer's Telegrams have reported an X-ray activity of \sgra{} that is still high in 2019, as observed by {\it Swift} and the ART-XC telescope.
We verified this result by analyzing the {\it Swift} data obtained in 2019.
We detected four flares, which is the largest number of flares observed during a single campaign.
The analysis of the flaring rate including these four flares shows that the high bright flaring rate persists.
However, the new {\it Chandra} and {\it XMM-Newton} observations performed during 2019 (becoming public in 2020) are required to detect X-ray flares with a lower flux compared to what can be detected with {\it Swift} and to obtain a better characterization of their duration.

Since 2014, the activity of \sgra{} thus increased in several wavelengths.
Additional multiwavelength data are required to conclude on the persistence of this increase and to obtain clues on the source of this unprecedented activity of the supermassive black hole.

\begin{acknowledgements}
We acknowledge support through an XMM PRODEX contract (Belspo), from the Fonds de la Recherche Scientifique (FRS/FNRS), as well as through an ARC grant for Concerted Research Actions financed by the French Community of Belgium (Wallonia-Brussels Federation). 
We thank the PIs that obtained since 2015 the X-ray observations of \sgra{} used in this work. 
This work made use of public data from the Swift data archive, and data supplied by the UK Swift Science Data Center at the University of Leicester. Swift is supported at Penn State University by NASA Contract NAS5-00136. 
This research has made use of the XRT Data Analysis Software (XRTDAS) developed under the responsibility of the ASI Science Data Center (ASDC), Italy. This work is also based on public data from the XMM-Newton project, which is an ESA Science Mission with instruments and contributions
\end{acknowledgements}

\begin{appendix}

\section{Observation log and X-ray flares detected from 2016 to 2018}
\begin{table*}
\caption[Observation log of the XMM-Newton observation in 2016--2018]{Observation log of public {\it XMM-Newton} observations in 2016--2018.}
\begin{center}
\resizebox{\textwidth}{!}{
\label{table:xmm}
\begin{tabular}{@{}rlllcrccccccc@{}}
\hline
\hline
\multicolumn{7}{c}{Observations} & \multicolumn{6}{c}{Flares}\\
\multicolumn{7}{c}{\rule[0.5ex]{47em}{0.55pt}} & \multicolumn{6}{c}{\rule[0.5ex]{33em}{0.55pt}}\\
\multicolumn{11}{c}{} & \multicolumn{2}{c}{Mean}\\
\multicolumn{11}{c}{} & \multicolumn{2}{c}{\rule[0.5ex]{16em}{0.55pt}}\\
\multicolumn{1}{c}{ObsID} & \multicolumn{1}{c}{PI} & \multicolumn{1}{c}{Start$\,$\tablefootmark{a}} & \multicolumn{1}{c}{End$\,$\tablefootmark{a}} & \multicolumn{1}{c}{Exposure$\,$\tablefootmark{b}}  & \multicolumn{1}{c}{Instrument}  & \multicolumn{1}{c}{Nonflaring level} & \multicolumn{1}{c}{number} & \multicolumn{1}{c}{Start} & \multicolumn{1}{c}{Stop} & \multicolumn{1}{c}{Duration} & \multicolumn{1}{c}{Count rate} & \multicolumn{1}{c}{Flux} \\
 & & \multicolumn{1}{c}{(UT)} & \multicolumn{1}{c}{(UT)} & \multicolumn{1}{c}{(ks)} & & \multicolumn{1}{c}{($\mathrm{count\ s^{-1}}$)} & & \multicolumn{1}{c}{(UT)} & \multicolumn{1}{c}{(UT)} & \multicolumn{1}{c}{(s)} & \multicolumn{1}{c}{($\mathrm{count\ s^{-1}}$)} &\multicolumn{1}{c}{($10^{-12}\ \mathrm{erg\ s^{-1}\ cm^{-2}}$)}\\
\hline  
0790180401& Schartel & 2016-02-26 16:20:13 & 2016-02-27 02:36:53 & 37.0 & EPIC/pn$\,$\tablefootmark{c} & $3.5\pm0.02$ & \dots & \dots\dots\dots & \dots\dots\dots & \dots\dots & \dots\dots\dots\dots& \dots\dots\dots\dots \\
\hline
\end{tabular}
}
\tablefoot{
\tablefoottext{a} {First and last good time intervals (GTI) start and stop in universal time (UT) reference.}
\tablefoottext{b} {The sum of the GTI.}
\tablefoottext{c} {The EPIC camera was in timing mode.}
}
\normalsize
\end{center}
\end{table*}

\begin{table*}
\caption[Observation log of the Chandra observations and the detected X-ray flares in 2016--2018]{Observation log of public {\it Chandra} observations and the detected X-ray flares in 2016--2018.}
\begin{center}
\resizebox{\textwidth}{!}{
\label{table:chandra1}
\begin{tabular}{@{}rlllcrccrrccc@{}}
\hline
\hline
\multicolumn{7}{c}{Observations} & \multicolumn{6}{c}{Flares}\\
\multicolumn{7}{c}{\rule[0.5ex]{47em}{0.55pt}} & \multicolumn{6}{c}{\rule[0.5ex]{36em}{0.55pt}}\\
\multicolumn{11}{c}{} & \multicolumn{2}{c}{Mean}\\
\multicolumn{11}{c}{} & \multicolumn{2}{c}{\rule[0.5ex]{16em}{0.55pt}}\\
\multicolumn{1}{c}{ObsID} & \multicolumn{1}{c}{PI} & \multicolumn{1}{c}{Start$\,$\tablefootmark{a}} & \multicolumn{1}{c}{End$\,$\tablefootmark{a}} & \multicolumn{1}{c}{Exposure$\,$\tablefootmark{b}}  & \multicolumn{1}{c}{Instrument}  & \multicolumn{1}{c}{Nonflaring level} & \multicolumn{1}{c}{number} & \multicolumn{1}{c}{Start} & \multicolumn{1}{c}{Stop} & \multicolumn{1}{c}{Duration} & \multicolumn{1}{c}{Count rate$\,$\tablefootmark{c}} & \multicolumn{1}{c}{Flux$\,$\tablefootmark{d}} \\
 & & \multicolumn{1}{c}{(UT)} & \multicolumn{1}{c}{(UT)} & \multicolumn{1}{c}{(ks)} & & \multicolumn{1}{c}{($\mathrm{count\ s^{-1}}$)} & & \multicolumn{1}{c}{(UT)} & \multicolumn{1}{c}{(UT)} & \multicolumn{1}{c}{(s)} & \multicolumn{1}{c}{($\mathrm{count\ s^{-1}}$)} &\multicolumn{1}{c}{($10^{-12}\ \mathrm{erg\ s^{-1}\ cm^{-2}}$)}\\
\hline  
18055& Garmire & 2016-02-13 08:59:23 & 2016-02-13 16:26:00 & 22.7 & ACIS-S & $0.0083 \pm 0.0009$ & 1 & 12:14:10 & 12:24:07 & 597 & $0.078 \pm 0.012$ & $11.6 \pm 1.80$ \\
\rule[0.5ex]{2.5em}{0.55pt} & \rule[0.5ex]{3em}{0.55pt} & \rule[0.5ex]{8.5em}{0.55pt} & \rule[0.5ex]{8.5em}{0.55pt} & \rule[0.5ex]{2.5em}{0.55pt} & \rule[0.5ex]{3em}{0.55pt} & \rule[0.5ex]{6.5em}{0.55pt} & 2 & 13:00:45 & 13:20:23 & 1178 & $0.067 \pm 0.008$ & $9.92 \pm 1.18$ \\
\rule[0.5ex]{2.5em}{0.55pt} & \rule[0.5ex]{3em}{0.55pt} & \rule[0.5ex]{8.5em}{0.55pt} & \rule[0.5ex]{8.5em}{0.55pt} & \rule[0.5ex]{2.5em}{0.55pt} & \rule[0.5ex]{3em}{0.55pt} & \rule[0.5ex]{6.5em}{0.55pt} & 3 & 15:30:17 & $>$16:26:00 & $>$2129 & $0.078 \pm 0.006$ & $11.6 \pm 0.89$ \\
18056 & Garmire & 2016-02-14 14:46:01 & 2016-02-14 21:44:19 & 21.8 & ACIS-S & $0.0090 \pm 0.0006$  & \dots & \dots\dots\dots & \dots\dots\dots & \dots\dots & \dots\dots\dots\dots& \dots\dots\dots\dots \\
18731 & Baganoff & 2016-07-12 18:23:59 & 2016-07-13 18:42:51 & 78.4 & ACIS-S & $0.0047 \pm 0.0003$ & 4 & 22:43:30 & 22:58:26 & 827 & $0.029 \pm 0.004$ & $4.29 \pm 0.59$ \\
18732 & Baganoff & 2016-07-18 12:01:38 & 2016-07-19 12:09:00 & 76.6 & ACIS-S & $0.0047 \pm 0.0003$ & 5 & 14:56:24 & 15:47:03 & 3039 & $0.017 \pm 0.003$ & $2.52 \pm 0.44$ \\
18057 & Garmire & 2016-10-08 19:07:12 & 2016-10-09 02:38:59 & 22.7 & ACIS-S & $0.0054 \pm 0.0005$ & \dots & \dots\dots\dots & \dots\dots\dots & \dots\dots & \dots\dots\dots\dots& \dots\dots\dots\dots \\
18058 & Garmire & 2016-10-14 10:47:43 & 2016-10-14 18:16:44 & 22.7 & ACIS-S & $0.0047 \pm 0.0004$ & \dots & \dots\dots\dots & \dots\dots\dots & \dots\dots & \dots\dots\dots\dots & \dots\dots\dots\dots\\
19726 & Garmire & 2017-04-06 03:47:13 & 2017-04-06 12:51:35 & 28.2 & ACIS-S & $0.0045 \pm 0.0004$ & \dots & \dots\dots\dots & \dots\dots\dots & \dots\dots & \dots\dots\dots\dots & \dots\dots\dots\dots\\
19727 & Garmire & 2017-04-07 04:57:18 & 2017-04-07 13:53:40 & 27.8 & ACIS-S & $0.0059 \pm 0.0004$ & \dots & \dots\dots\dots & \dots\dots\dots & \dots\dots & \dots\dots\dots\dots& \dots\dots\dots\dots \\
20041 & Garmire & 2017-04-11 03:51:22 & 2017-04-11 13:56:48 & 30.9 & ACIS-S & $0.0066 \pm 0.0006$ & 6 & 08:23:04 & 09:56:59 & 2635 & $0.085 \pm 0.006$ & $12.6 \pm 0.89$ \\
20040 & Garmire & 2017-04-12 05:18:22 & 2017-04-12 14:15:52 & 27.5 & ACIS-S & $0.0051 \pm 0.0004$ & \dots & \dots\dots\dots & \dots\dots\dots & \dots\dots & \dots\dots\dots\dots & \dots\dots\dots\dots\\
19703 & Baganoff & 2017-07-15 22:36:07 & 2017-07-17 00:01:34 & 81.0 & ACIS-S & $0.0046 \pm 0.0003$ & 7 & $<$22:36:07 & 24:30:18 & $>$5736 & $0.008 \pm 0.002$ & $1.18 \pm 0.22$ \\
\rule[0.5ex]{2.5em}{0.55pt} & \rule[0.5ex]{3em}{0.55pt} & \rule[0.5ex]{8.5em}{0.55pt} & \rule[0.5ex]{8.5em}{0.55pt} & \rule[0.5ex]{2.5em}{0.55pt} & \rule[0.5ex]{3em}{0.55pt} & \rule[0.5ex]{6.5em}{0.55pt} & 8 & 37:10:52 & 37:29:56 & 1144 & $0.029 \pm 0.005$ & $4.29 \pm 0.74$ \\
19704 & Baganoff & 2017-07-25 22:57:27 & 2017-07-26 23:28:30 & 78.4 & ACIS-S & $0.0050 \pm 0.0002$ & \dots & \dots\dots\dots & \dots\dots\dots & \dots\dots & \dots\dots\dots\dots & \dots\dots\dots\dots\\
20344 & Neilsen & 2018-04-20 03:17:44 & 2018-04-20 12:59:33 & 29.1 & ACIS-S & $0.0053 \pm 0.0004$ & \dots & \dots\dots\dots & \dots\dots\dots & \dots\dots & \dots\dots\dots\dots& \dots\dots\dots\dots \\
20345 & Neilsen & 2018-04-22 03:31:16 & 2018-04-22 12:57:15 & 28.5 & ACIS-S & $0.0037 \pm 0.0003$ & \dots & \dots\dots\dots & \dots\dots\dots & \dots\dots & \dots\dots\dots\dots& \dots\dots\dots\dots \\
20346 & Neilsen & 2018-04-24 03:33:43 & 2018-04-24 13:21:29 & 29.0 & ACIS-S & $0.0042 \pm 0.0004$ & 9 & 05:00:57 & 05:53:52 & 3140 & $0.124 \pm 0.006$ & $18.4 \pm 0.89$ \\
20347 & Neilsen & 2018-04-25 03:37:23 & 2018-04-25 14:13:22 & 32.8 & ACIS-S & $0.0048 \pm 0.0004$ & \dots & \dots\dots\dots & \dots\dots\dots & \dots\dots & \dots\dots\dots\dots& \dots\dots\dots\dots \\
\hline
\end{tabular}
}
\tablefoot{
\tablefoottext{a} {First and last good time intervals (GTI) start and stop in universal time (UT) referential.}
\tablefoottext{b} {The sum of the GTI.}
\tablefoottext{c} {The flare mean count rates are background subtracted.}
\tablefoottext{d} {Mean unabsorbed flux between 2 and 10~keV determined for $N_\mathrm{H}=14.3\times 10^{22}\ \mathrm{cm^{-2}}$ and $\Gamma=2$ with the pile-up taken into account.}
}
\normalsize
\end{center}
\end{table*}

\begin{table*}
\caption[Observation log of the Swift observations and the detected X-ray flares in 2016--2018]{Observation log of public {\it Swift} observations and the detected X-ray flares in 2016--2018.}
\begin{center}
\resizebox{\textwidth}{!}{
\label{table:swift}
\begin{tabular}{@{}llccccccccc@{}}
\hline
\hline
\multicolumn{5}{c}{Observations} & \multicolumn{6}{c}{Flares}\\
\multicolumn{5}{c}{\rule[0.5ex]{40em}{0.55pt}} & \multicolumn{6}{c}{\rule[0.5ex]{44em}{0.55pt}}\\
\multicolumn{9}{c}{} & \multicolumn{2}{c}{Mean}\\
\multicolumn{9}{c}{} & \multicolumn{2}{c}{\rule[0.5ex]{18em}{0.55pt}}\\
\multicolumn{1}{c}{First} & \multicolumn{1}{c}{Last} & \multicolumn{1}{c}{Number} & \multicolumn{1}{c}{Total exposure}  & \multicolumn{1}{c}{Nonflaring level$\,$\tablefootmark{a}} & \multicolumn{1}{c}{number} & \multicolumn{1}{c}{Start} & \multicolumn{1}{c}{Stop} & \multicolumn{1}{c}{Duration} & \multicolumn{1}{c}{Count rate$\,$\tablefootmark{b}} & \multicolumn{1}{c}{Flux$\,$\tablefootmark{c}} \\
 \multicolumn{1}{c}{(UT)} & \multicolumn{1}{c}{(UT)} & & \multicolumn{1}{c}{(ks)} & \multicolumn{1}{c}{($\mathrm{count\ s^{-1}}$)} & & \multicolumn{1}{c}{(UT)} & \multicolumn{1}{c}{(UT)} & \multicolumn{1}{c}{(s)} & \multicolumn{1}{c}{($\mathrm{count\ s^{-1}}$)} &\multicolumn{1}{c}{($10^{-12}\ \mathrm{erg\ s^{-1}\ cm^{-2}}$)}\\
\hline  
2016-02-06 20:58:58 &  2016-07-30 13:11:57 &  153 &  142.6 &  $0.377 \pm 0.089$ &  1 &  2016-03-24 19:29:41 &  2016-03-24 19:43:54 &  853 &  $0.483 \pm 0.094$ &  $141.8 \pm 27.6$ \\
\rule[0.5ex]{8.5em}{0.55pt} & \rule[0.5ex]{8.5em}{0.55pt} & \rule[0.5ex]{1.5em}{0.55pt} & \rule[0.5ex]{2.5em}{0.55pt} & $0.238 \pm 0.065$ & 2 &  2016-05-05 00:36:01 &  2016-05-05 00:43:53 &  472 &  $0.300 \pm 0.073$ &  $88.1 \pm 21.4$ \\
\rule[0.5ex]{8.5em}{0.55pt} & \rule[0.5ex]{8.5em}{0.55pt} & \rule[0.5ex]{1.5em}{0.55pt} & \rule[0.5ex]{2.5em}{0.55pt} & $0.061 \pm 0.024$ & 3\,\tablefootmark{d} &  2016-05-19 10:23:39 &  2016-05-19 10:39:40 &  961 &  $0.116 \pm 0.0.028$ &  $34.1 \pm 8.2$ \\
\rule[0.5ex]{8.5em}{0.55pt} & \rule[0.5ex]{8.5em}{0.55pt} & \rule[0.5ex]{1.5em}{0.55pt} & \rule[0.5ex]{2.5em}{0.55pt} & $0.068 \pm 0.015$ & 4\,\tablefootmark{d} &  2016-05-30 09:42:54 &  2016-05-30 09:58:54 &  960 &  $0.080 \pm 0.0.019$ &  $23.5 \pm 5.6$ \\
2016-08-01 17:31:57 &  2016-11-01 07:06:58 &  78 &  75.4 &  $0.028 \pm 0.0001$ &  \dots &  \dots\dots\dots\dots\dots\dots\dots &  \dots\dots\dots\dots\dots\dots\dots &  \dots\dots &  \dots\dots\dots\dots\dots &  \dots\dots\dots\dots \\
2017-02-02 23:31:57 &  2017-11-02 00:26:57 &  291 &  248.5 &  $0.026 \pm 0.0003$ &  5 &  2017-06-12 15:59:41 &  2017-06-12 16:20:52 &  1271 &  $0.096 \pm 0.010$ &  $28.2 \pm 2.94$ \\ 
2018-02-02 17:58:56 &  2018-11-02 04:27:57 &  233 &  222.7 &  $0.025 \pm 0.0003$ &  6 &  2018-02-17 00:45:29 &  2018-02-17 01:01:52 &  983 & $0.069 \pm 0.010$ &  $20.2 \pm 2.94$ \\
\rule[0.5ex]{8.5em}{0.55pt} & \rule[0.5ex]{8.5em}{0.55pt} & \rule[0.5ex]{1.5em}{0.55pt} & \rule[0.5ex]{2.5em}{0.55pt} & \rule[0.5ex]{6em}{0.55pt} & 7 &  2018-08-22 19:17:33 &  2018-08-22 19:29:53 &  740 &  $0.083 \pm 0.012$ &  $24.4 \pm 3.52$ \\
\hline
\end{tabular}
}
\tablefoot{
\tablefoottext{a} {In the first part of 2016, it was impossible to determine a unique nonflaring level because of the activity of X-ray transients. We show the four nonflaring levels, associated with the four flares, computed as the mean count rate produced by the transients.}
\tablefoottext{b} {The flare mean count rates are background subtracted.}
\tablefoottext{c} {Mean unabsorbed flux between 2 and 10~keV determined for $N_\mathrm{H}=14.3\times 10^{22}\ \mathrm{cm^{-2}}$ and $\Gamma=2$.}
\tablefoottext{d} {These flares cannot be unequivocally attributed to \sgra{}.
They are therefore ignored in the study of the flaring rate.}
}
\normalsize
\end{center}
\end{table*}

\section{Light curves of the X-ray flares detected from 2016 to 2018}
\begin{figure*}[t]
\centering
\includegraphics[trim=0.cm 0.cm 0cm 0cm,clip,width=5.cm,angle=90]{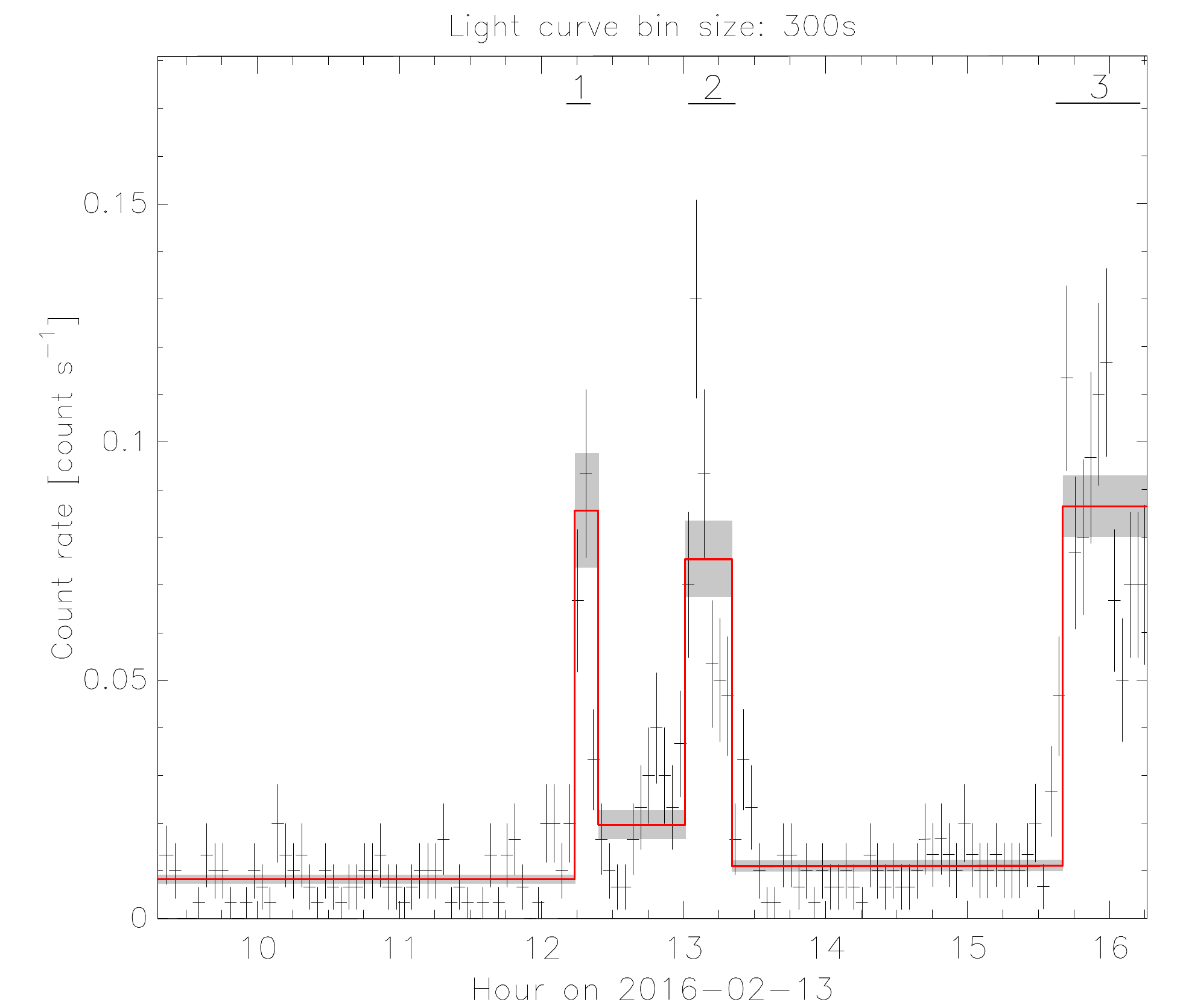}
\includegraphics[trim=0.cm 0.cm 0cm 0cm,clip,width=5.cm,angle=90]{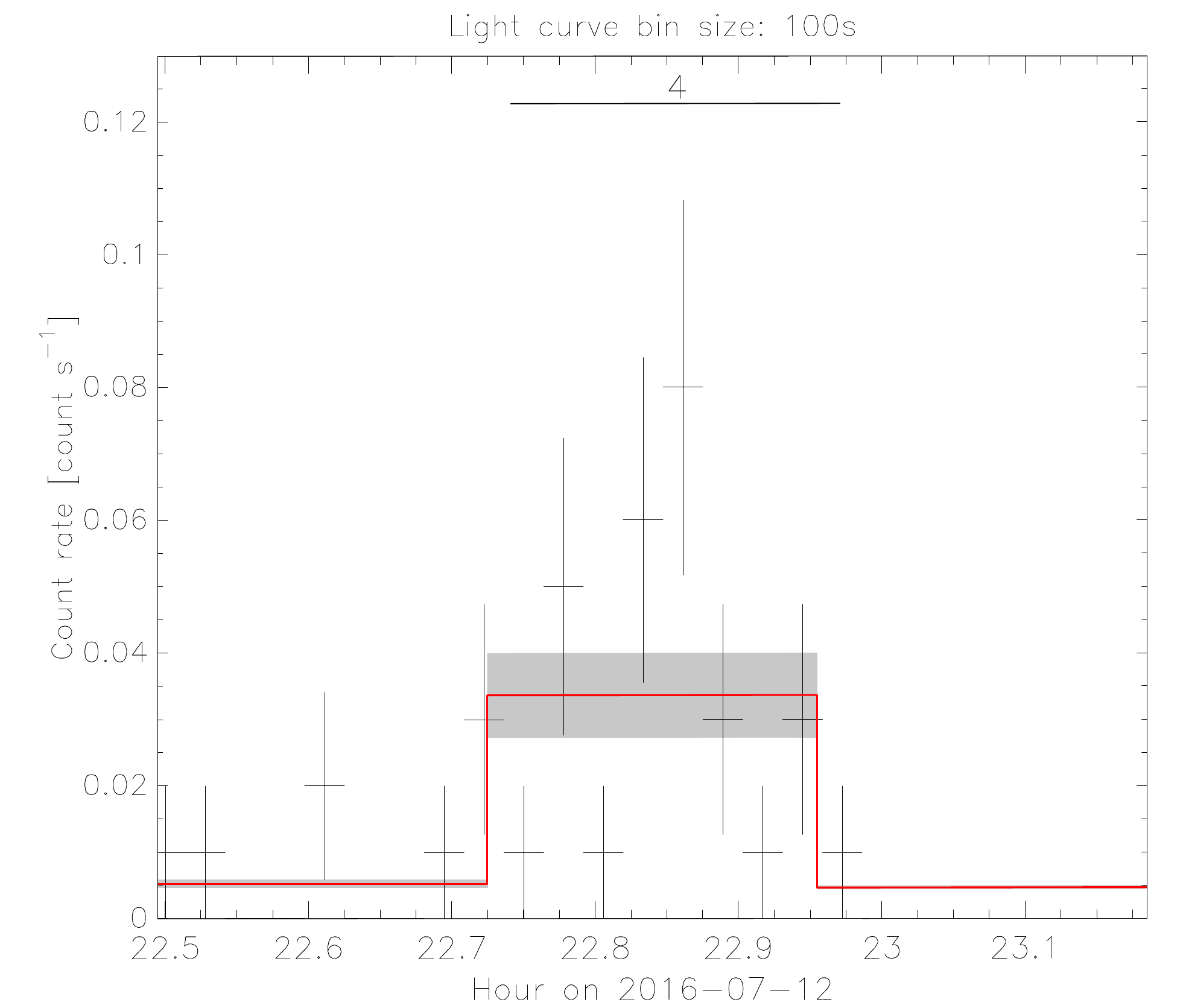}
\includegraphics[trim=0.cm 0.cm 0cm 0cm,clip,width=5.cm,angle=90]{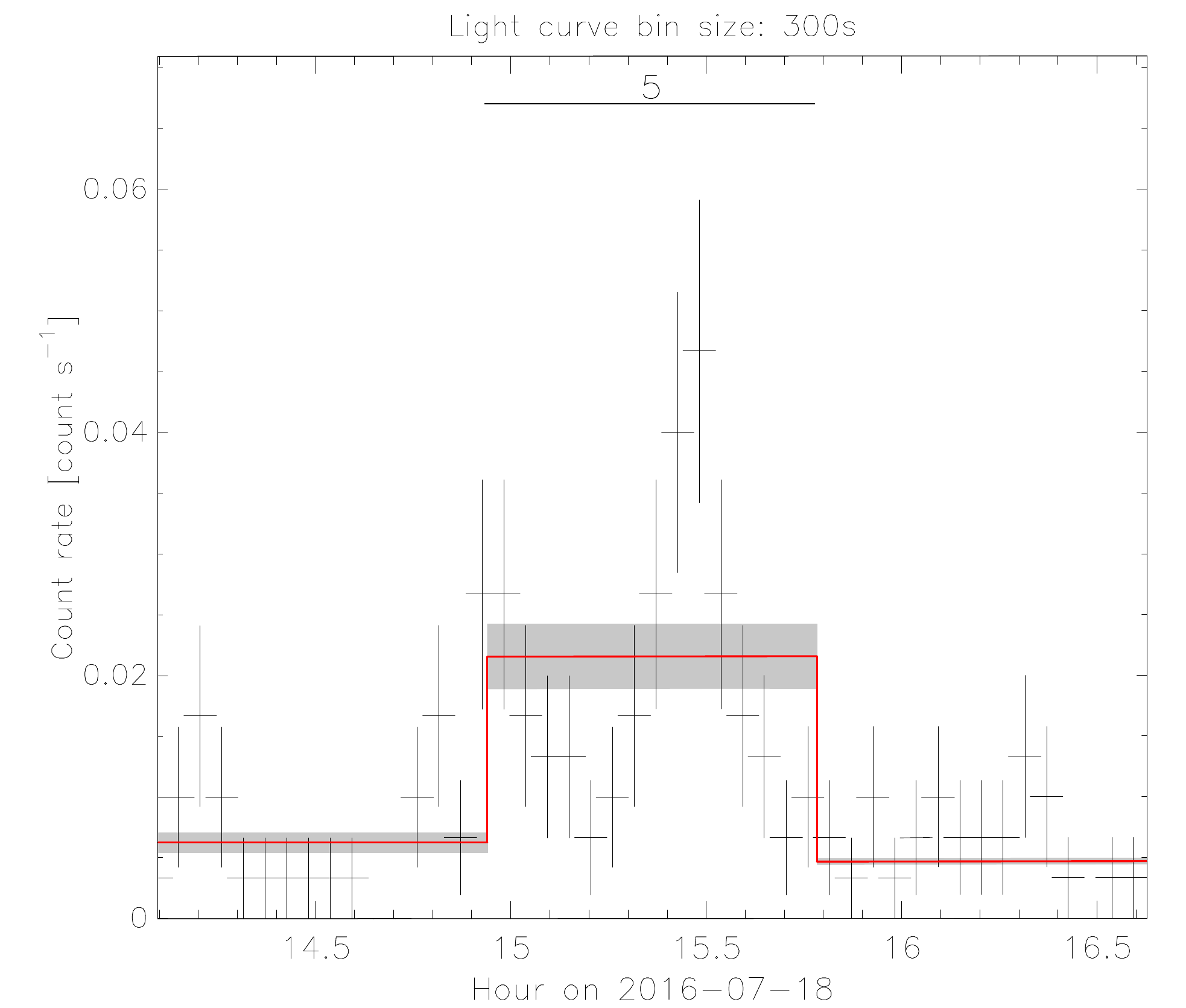}\\
\includegraphics[trim=0.cm 0.cm 0cm 0cm,clip,width=5.cm,angle=90]{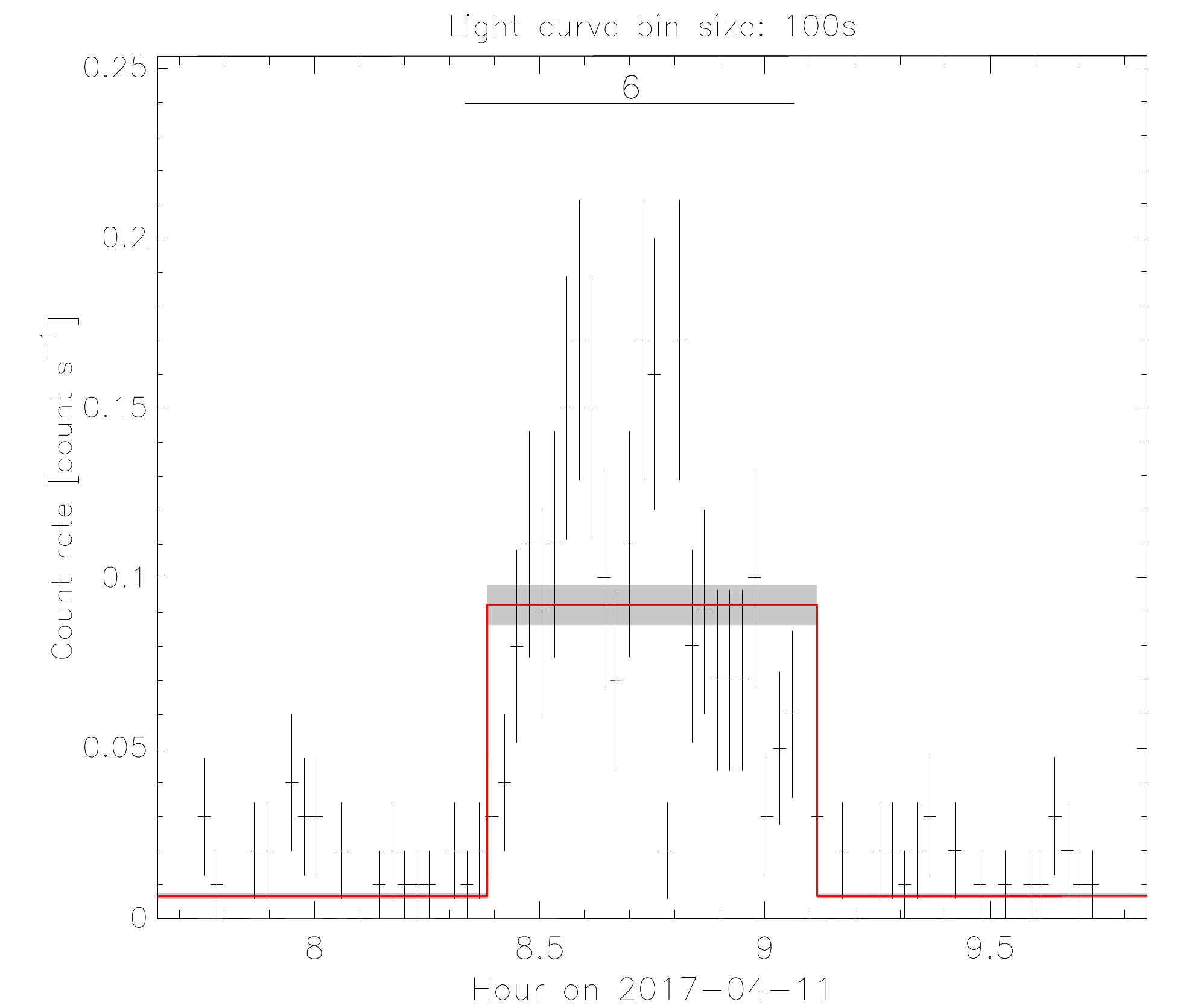}
\includegraphics[trim=0.cm 0.cm 0cm 0cm,clip,width=5.cm,angle=90]{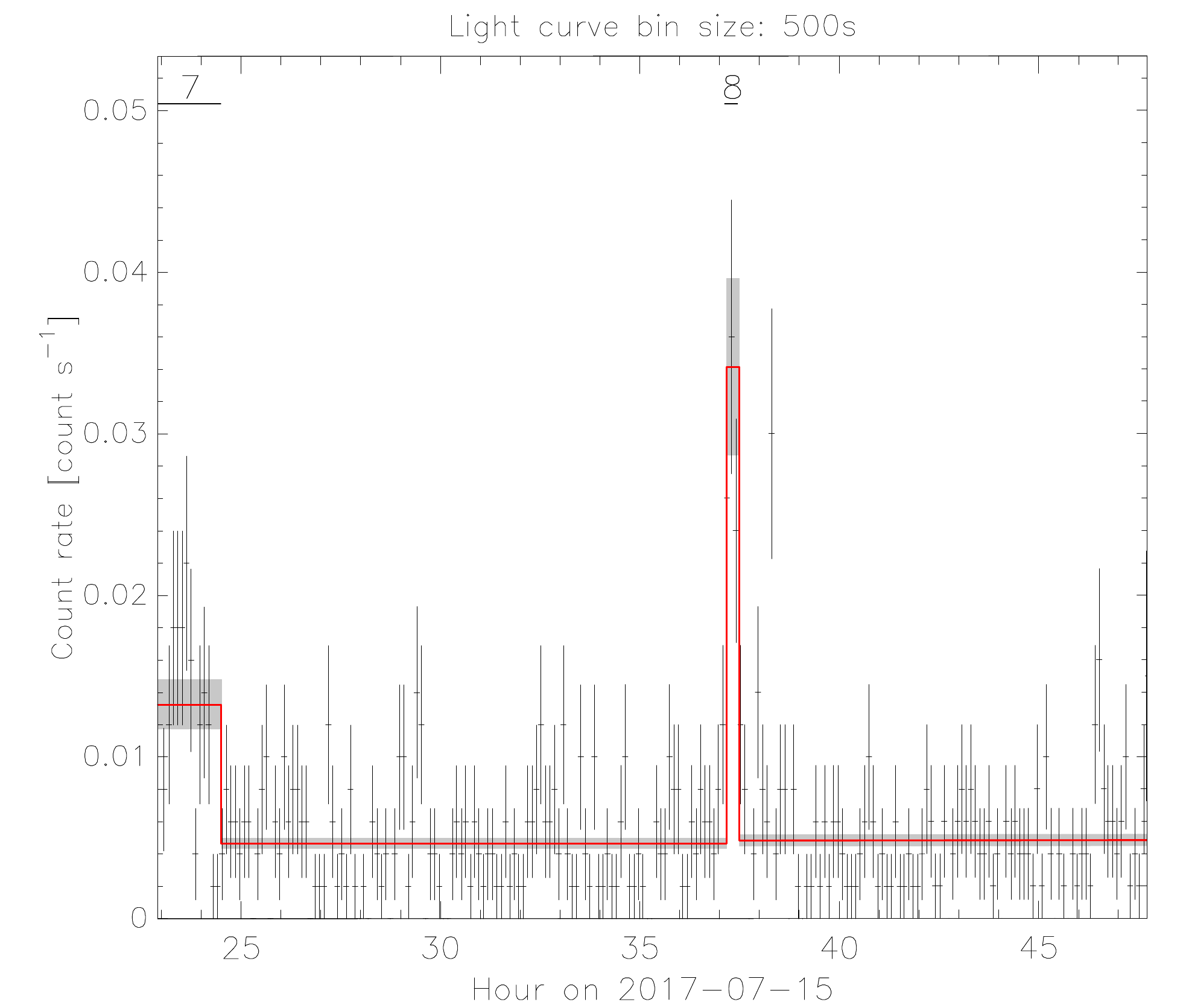}
\includegraphics[trim=0.cm 0.cm 0cm 0cm,clip,width=5.cm,angle=90]{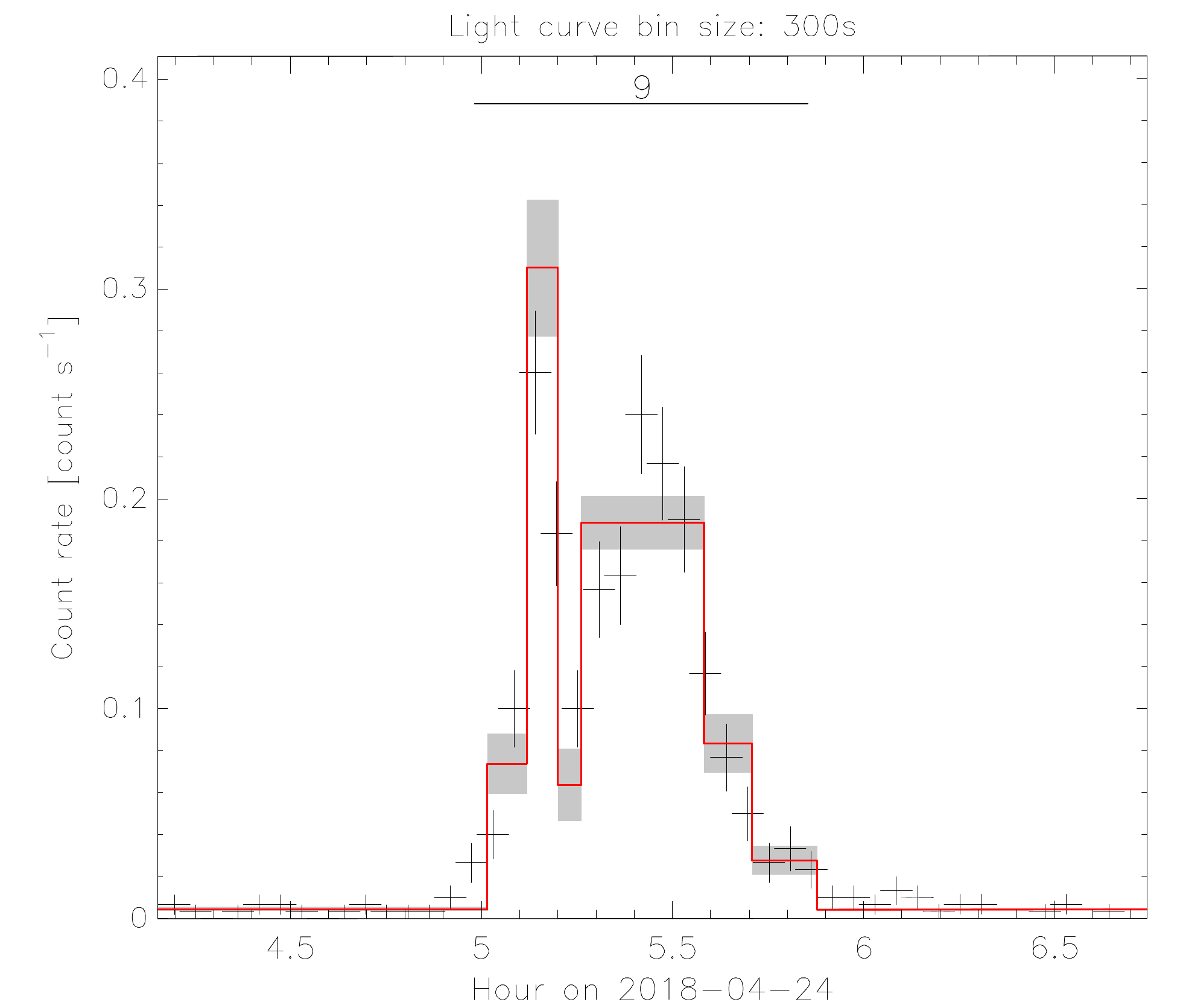}
\caption[Light curves of the X-ray flares.]{X-ray flares detected using the Bayesian blocks algorithm in 2016--2018.
All these flares were observed during {\it Chandra} observations.
The black crosses are the count rates and their error bars. 
The bin sizes of the light curves are reported at the top of each figure. 
The red lines are the Bayesian blocks and their errors are horizontal gray rectangles.
Each flare is labeled with the index corresponding to the flare number in Table~\ref{table:chandra1}.
The horizontal line is the flare duration.}
\label{fig:light_curves}
\end{figure*}

\section{Spectral analysis of the X-ray flares observed by {\it Chandra} on 2016 February\ 13}
\label{spectre}
\begin{figure*}[t]
\centering
\includegraphics[trim=1.cm 0.cm 0cm 0cm,clip,width=4.1cm,angle=-90]{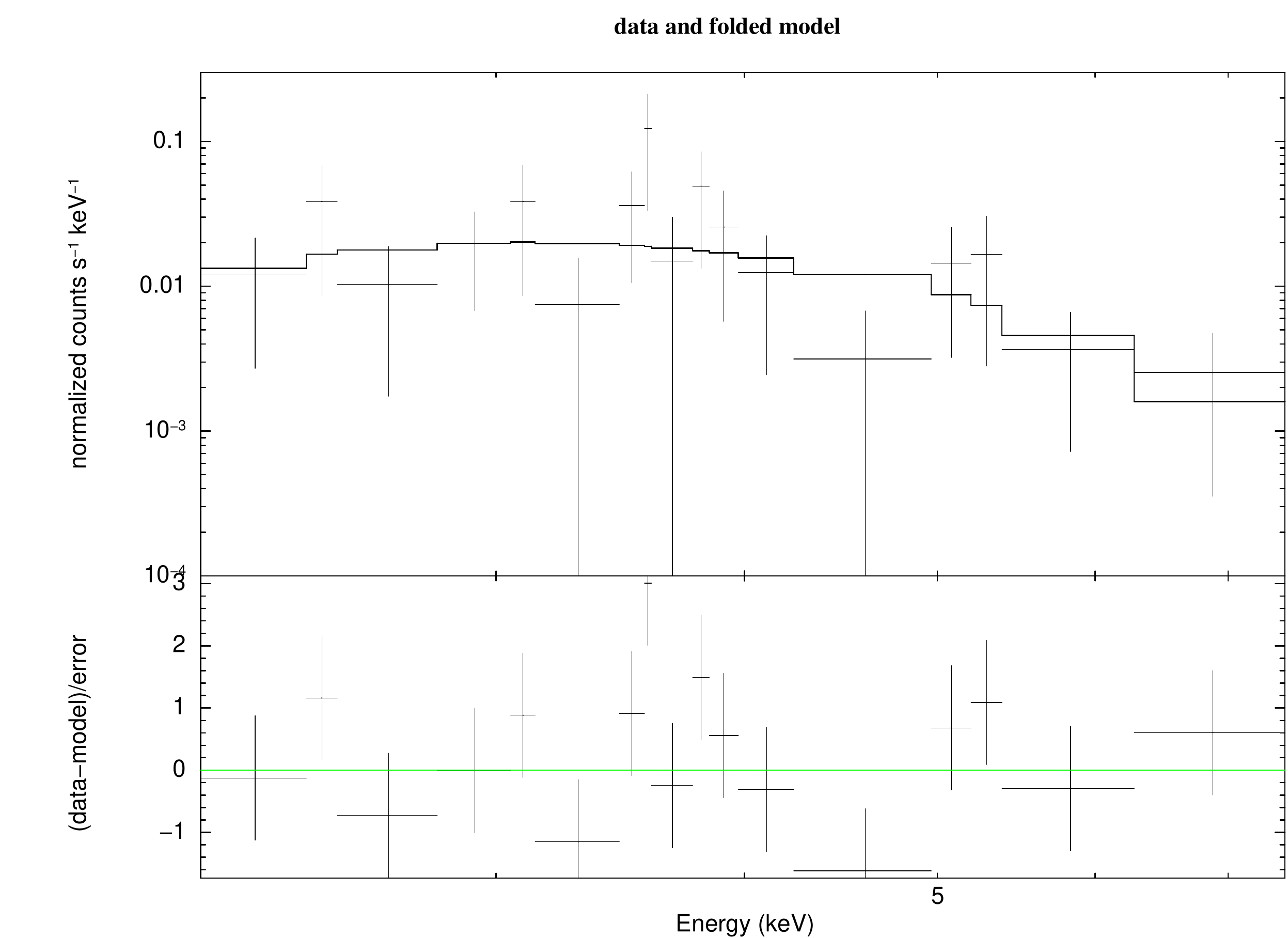}
\includegraphics[trim=1.cm 0.cm 0cm 0cm,clip,width=4.1cm,angle=-90]{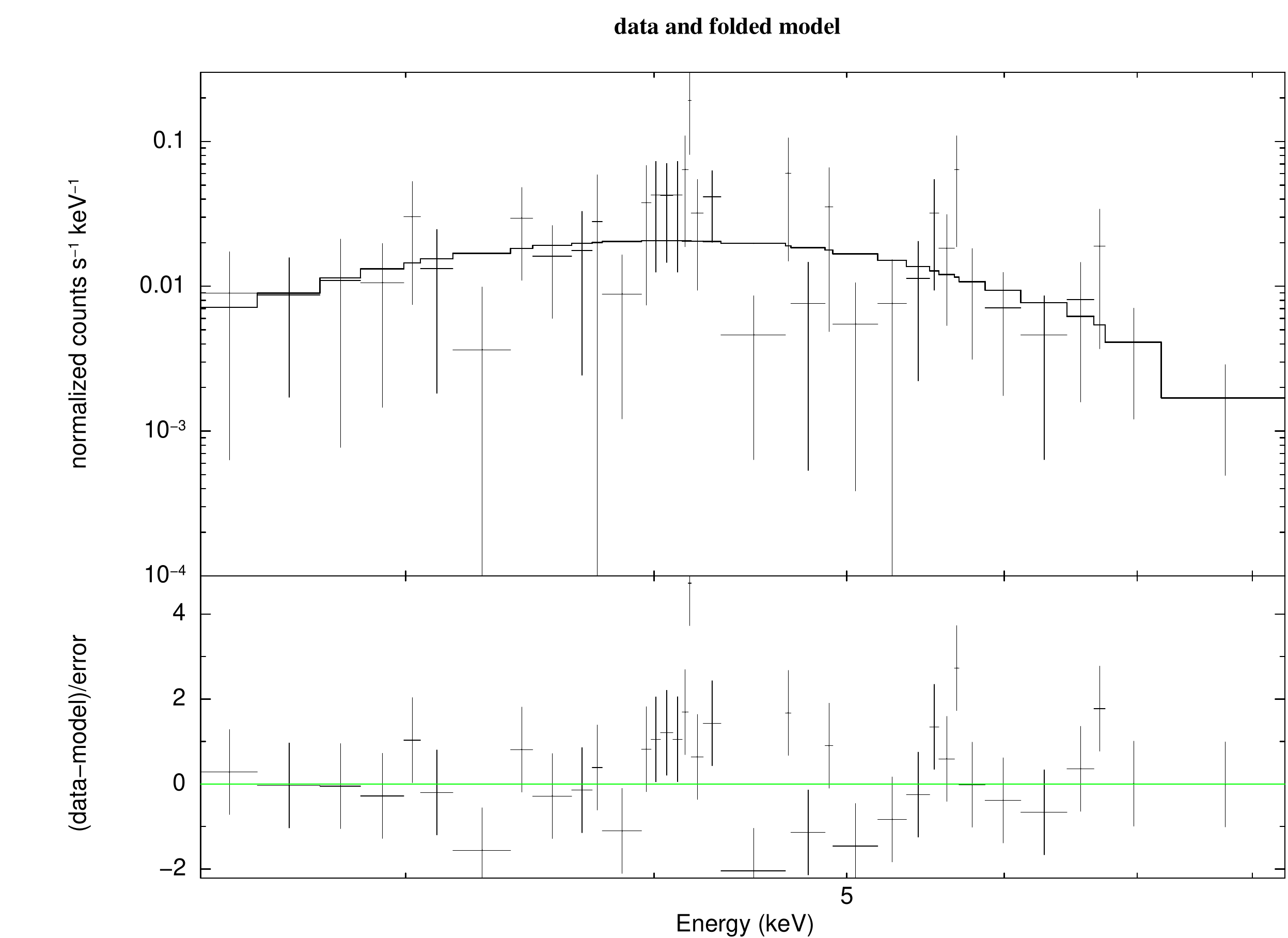}
\includegraphics[trim=1.cm 0.cm 0cm 0cm,clip,width=4.1cm,angle=-90]{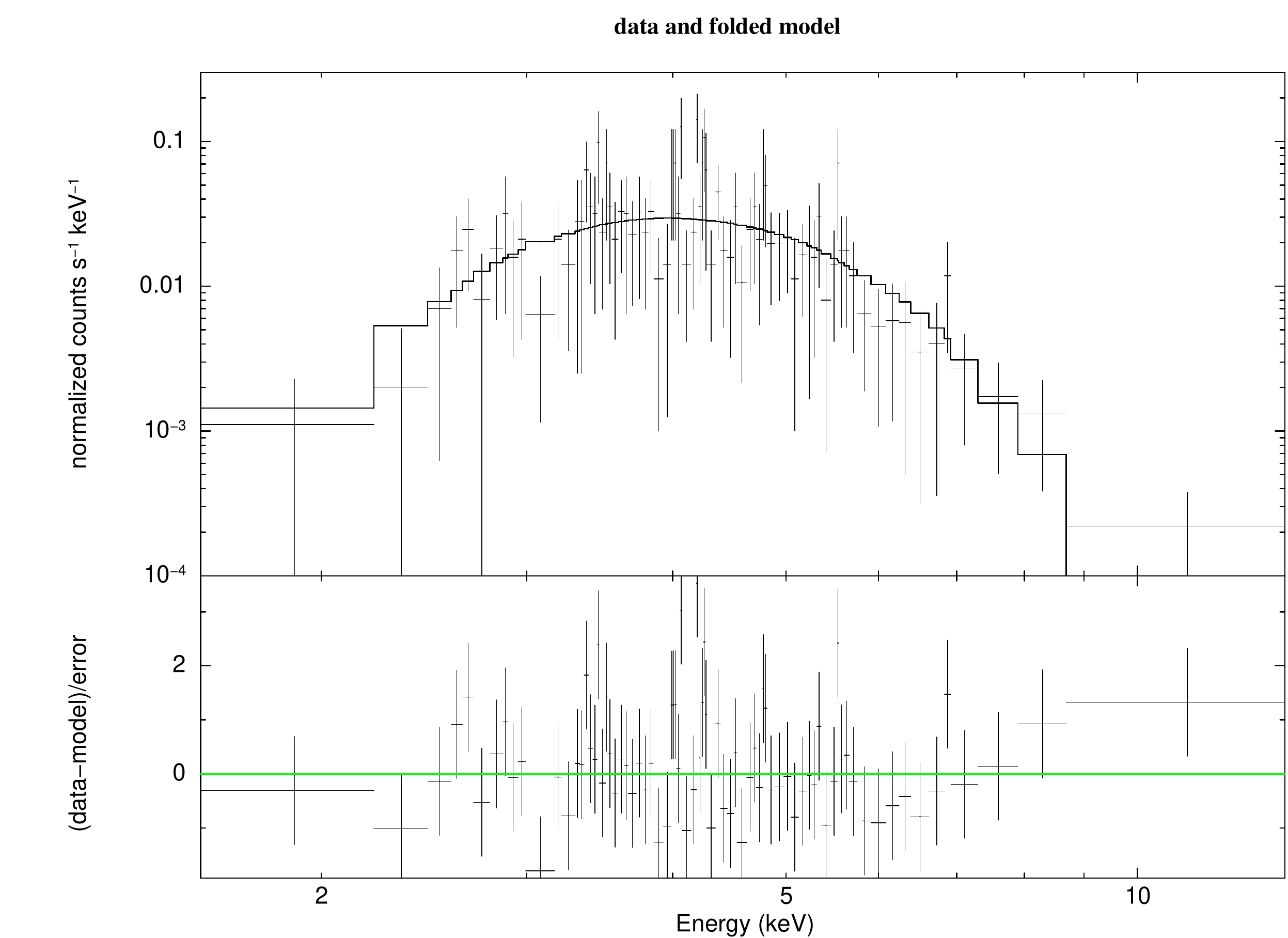}
\caption[Spectra of the X-ray flares observed on 2016 Feb.\ 13.]{Spectra of X-ray flares 1 to 3 (left to right) observed with {\it Chandra}.
The vertical bars are the 1$\sigma$ error in the count rate and the horizontal bars show the spectral bin energies. 
The events have been grouped with a minimum of two counts per bin.
{\it Top panels:} Best-fit is shown by the continuous solid line. 
{\it Bottom panels:} Residuals in units of $\sigma$.}
\label{fig:spectra}
\end{figure*}

We analyzed the X-ray spectra of the flares observed by {\it Chandra} on 2016 February\ 13 to verify that these emissions come from \sgra{} and not from the two transients.
We extracted the spectra, ancillary files, and response matrix files with the CIAO script \texttt{specextract}.
The spectra of the three flares were extracted from the flaring time intervals determined with the two-step Bayesian blocks algorithm.
The background spectrum (which is the same for the three flares) was extracted from the nonflaring time intervals.
Because of the low number of counts in the net spectra (source - background), we grouped them with a minimum of two counts per bin with \texttt{grppha} to fit them using the Cash statistic \citep{cash79} in \texttt{XSPEC}.
We fit the spectra with a power law model using {\it pegpwrlw}.
The power law was absorbed using {\it TBnew} \citep{wilms00} and dust-scattered with {\it dustscat} \citep{predehl95}.
Owing to the low fluxes of the flares, the pile-up, which may affect luminous sources observed by {\it Chandra}, was negligible.
Three spectral parameters were therefore estimated and compared to the typical values of the \sgra{} flares: the hydrogen column density ($N_\mathrm{H}$), the power law index ($\Gamma$), and the mean unabsorbed flux in 2--10\,keV ($F_\mathrm{2-10keV}$).
Figure~\ref{fig:spectra} and Table~\ref{tab:spectra} show the results of the fit.
The best-fit values of the hydrogen column densities and spectral indexes are similar to the typical values of $N_\mathrm{H}= 14.3 \times 10^{22}\,\mathrm{cm^{-2}}$ and $\Gamma = 2$ determined from the largest X-ray flares of \sgra{} \citep{porquet03,nowak12}.

\begin{table}
\caption[Best-fit values of the spectral parameters]{Best-fit values of the spectral parameters of the flares on 2016 February\ 13.}
\begin{center}
\resizebox{0.49\textwidth}{!}{
\label{tab:spectra}
\begin{tabular}{@{}ccccc@{}}
\hline
\hline
Flare & $N_\mathrm{H}$ & $\Gamma$ & $F_\mathrm{2-10keV}$ & cstat (dof)\\
(number) & ($10^{22}\,\mathrm{cm^{-2}}$) & & ($10^{-12}\,\mathrm{erg\,s^{-1}\,cm^{-2}}$) & \\
\hline  
1 & $8.2^{+17.2}_{-8.2}$ & $2.8^{+4.2}_{-1.5}$ & $3.3^{+44.5}_{-1.4}$ & 13.9 (14) \\
2 & $10.9^{+14.3}_{-10.9}$ & $1.4^{+2.2}_{-1.4}$ & $5.7^{+11.0}_{-2.0}$ & 42.5 (34) \\
3 & $18.0^{+7.7}_{-6.0}$ & $2.8^{+1.3}_{-1.1}$ & $11.4^{+14.4}_{-4.3}$ & 51.6 (76) \\ 
\hline
\end{tabular}
}
\end{center}
\end{table}

The spectrum of the {\it Chandra} observation of \transquinze{} on 2016 February\ 13 has been analyzed by \citet{corrales17}.
Because of the very high luminosity of the source, the authors extracted the readout streak spectra, which minimizes the contamination by the pile-up and dust-scattering halo.
They tested three absorbed spectral models: a power law, a one-temperature blackbody, and a power law plus one-temperature blackbody.
The three models satisfy the description of the emission of \transquinze.
The more accurate models are the absorbed power law with $N_\mathrm{H}= 14.9^{+0.9}_{-0.8} \times 10^{22}\,\mathrm{cm^{-2}}$ and $\Gamma = 3.9\pm0.2$ and the absorbed power law plus one-temperature blackbody with $N_\mathrm{H}= 11.6^{+1.0}_{-0.4} \times 10^{22}\,\mathrm{cm^{-2}}$, $\Gamma = -0.03^{+1.34}_{-0.45}$ , and $\kappa T=0.68^{+0.05}_{-0.06}\,$keV.
However, we cannot conclude on the consistency of their and our spectral parameters because our error bars are very large.
We therefore directly applied their two best spectral models on the spectra of {\it Chandra} flare 3, which has the largest number of counts, leading to the smallest error bars.
We first applied the absorbed power law model with $N_\mathrm{H}$ fixed to $14.9 \times 10^{22}\,\mathrm{cm^{-2}}$ and let the spectral index free.
The resulting spectral index is $\Gamma = 2.3\pm0.5,$ which is more than $3\sigma$ lower than the value determined by \citet{corrales17}.
We then tested the absorbed power law plus one-temperature blackbody with $N_\mathrm{H}$ fixed to $11.6\times 10^{22}\,\mathrm{cm^{-2}}$.
However, it was not possible to converge to a satisfying value of $\kappa T$.
We therefore fixed $\Gamma = -0.03$ and $\kappa T=0.68\,$keV and let only the normalization parameters of the power law and blackbody component vary.
Once again, the value of the normalization of the backbody component is very low (about $10^{-7}$) and consistent with zero within the error bars.
Moreover, the high-energy emission (above 5\,keV) created by this model is too high compared to the spectrum of flare 3.

These tests thus show that at the spectral characteristics of at least flare 3 are inconsistent with the emission from \transquinze.
The first two flares are too faint for a stringent conclusion.
However, examining the light curve extracted from a disk with radius $1\farcs25$ centered on \transquinze, we did not observe any large variation by a factor of at least four during several hundred seconds, which may explain flares 1 to 3.
The three flares observed by {\it Chandra} in the \sgra{} light curve are therefore very likely emitted by the SMBH and can thus be taken into account in the flaring rate study.

\section{Light curves of the four 2016 \textit{Swift} flares}
\label{swift_flares}
\begin{figure*}[t]
\centering
\includegraphics[trim=0.cm 0.cm 0cm 0cm,clip,width=5.7cm,angle=90]{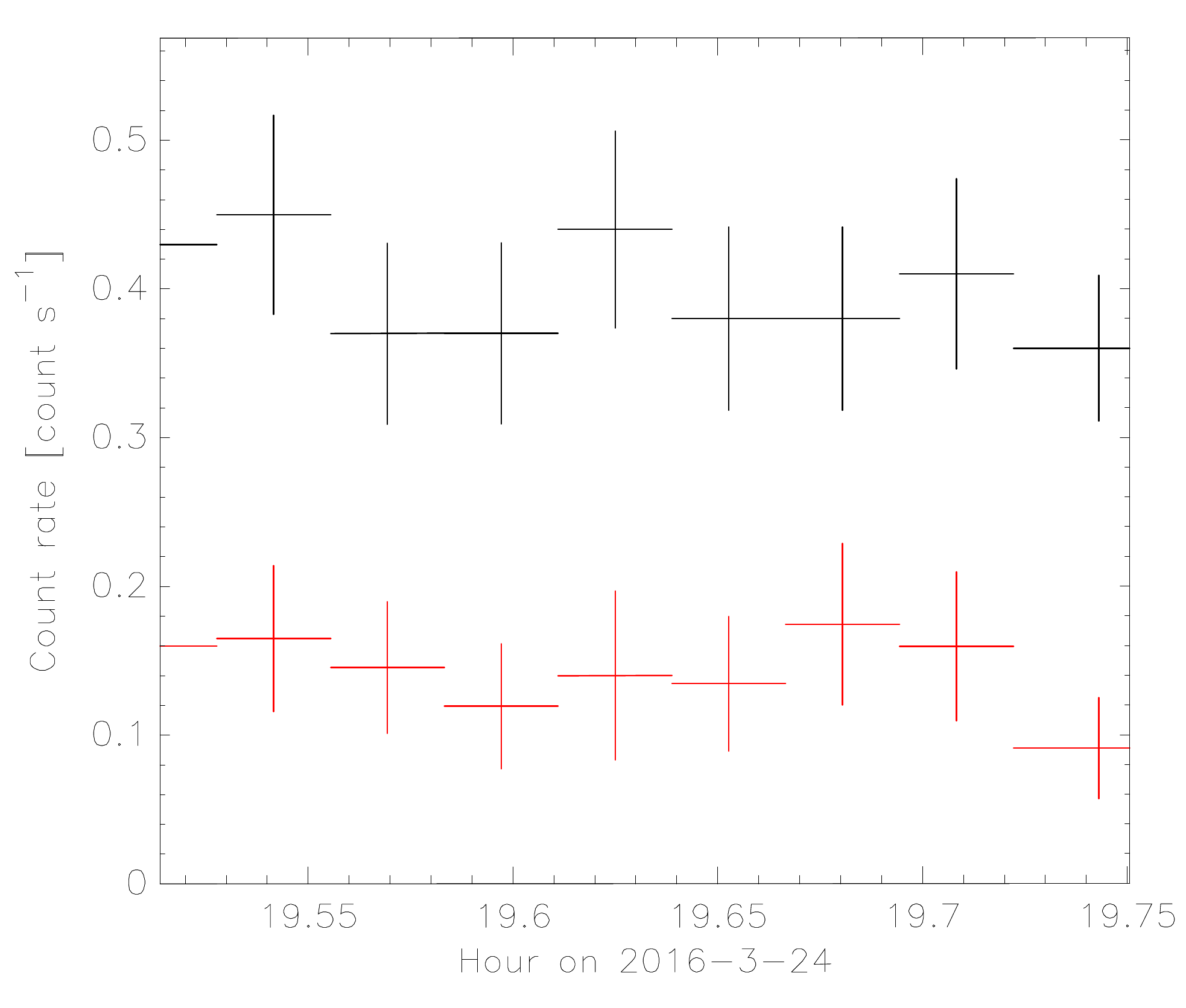}
\includegraphics[trim=0.cm 0.cm 0cm 0cm,clip,width=5.7cm,angle=90]{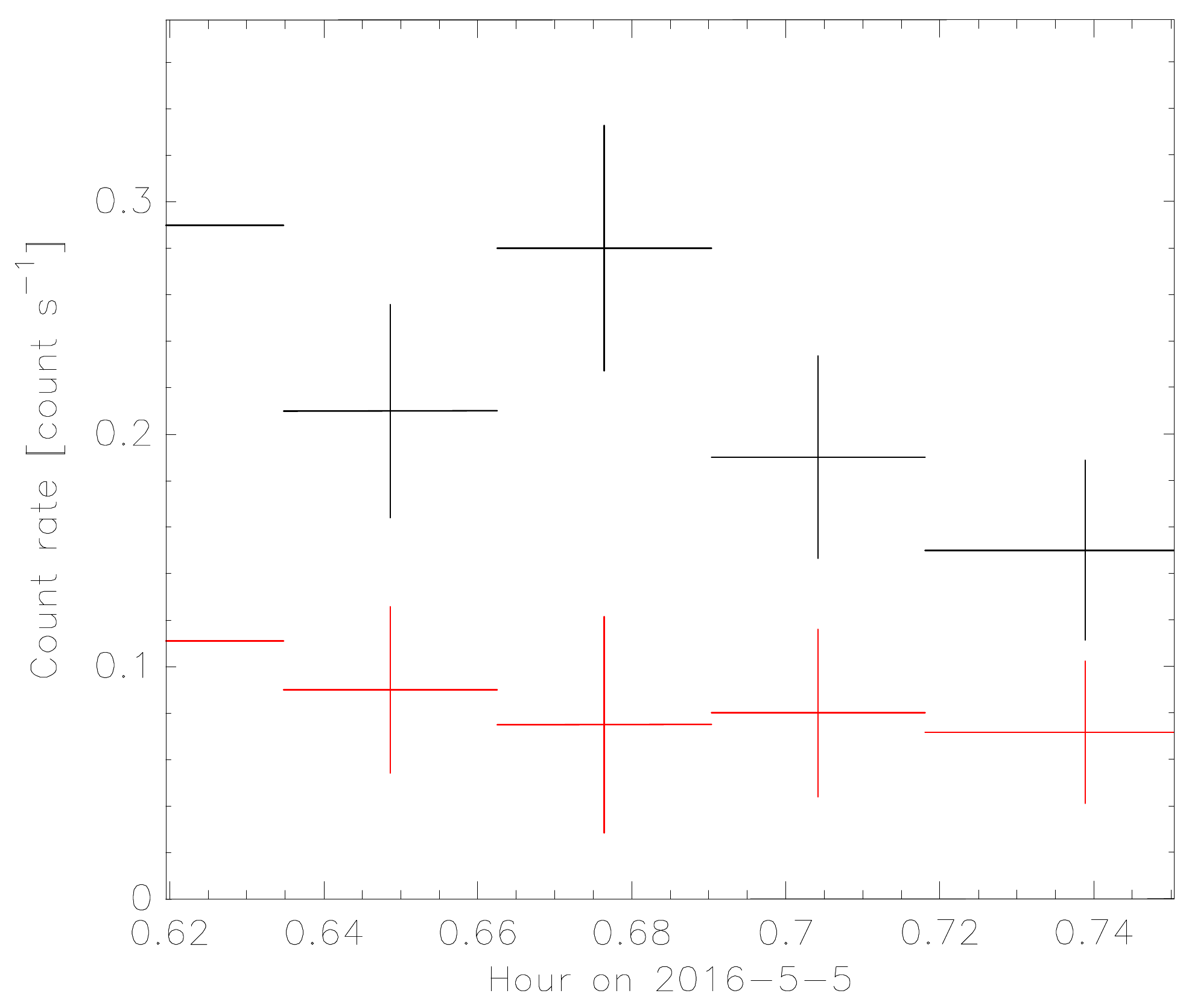}
\includegraphics[trim=0.cm 0.cm 0cm 0cm,clip,width=5.7cm,angle=90]{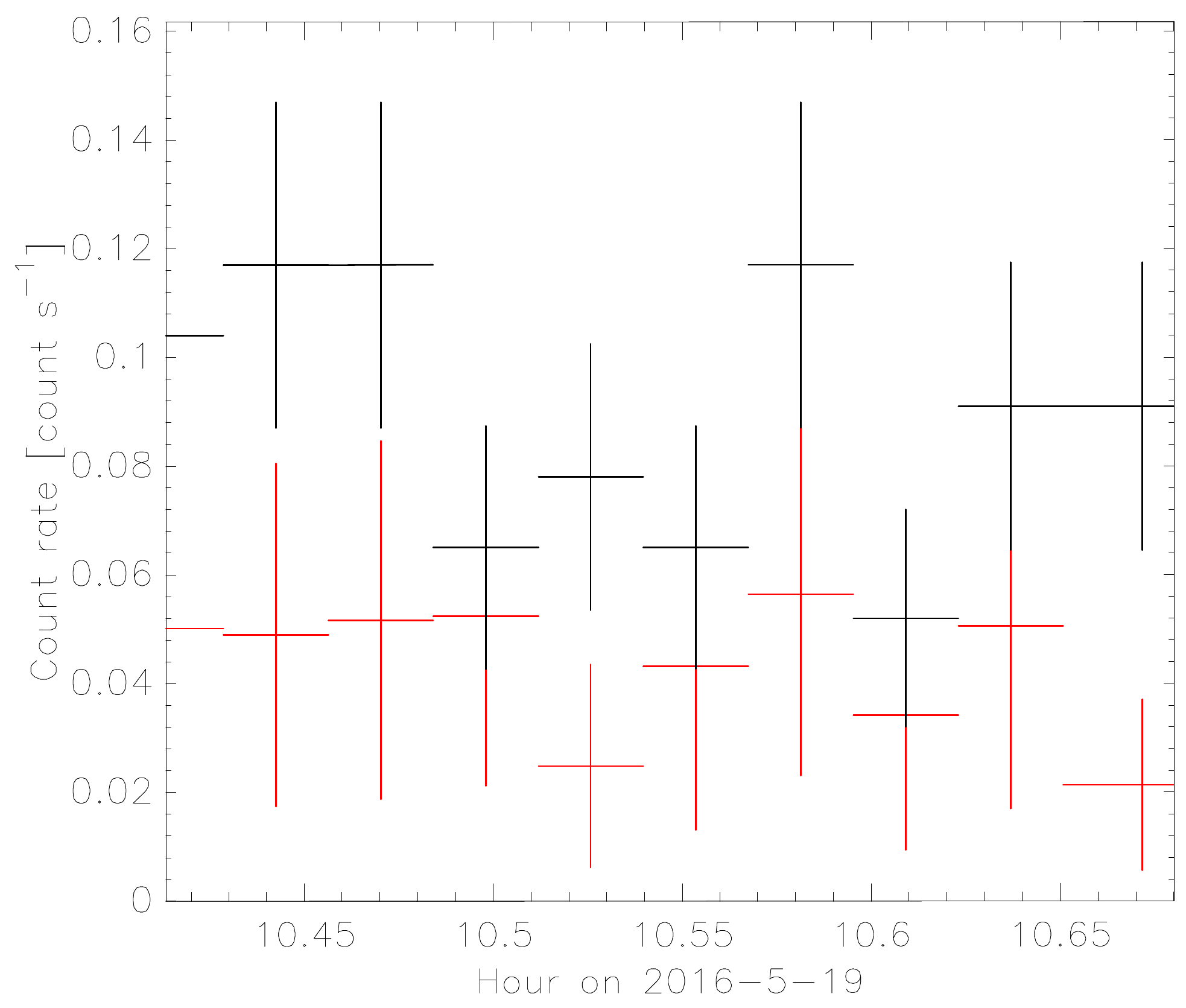}
\includegraphics[trim=0.cm 0.cm 0cm 0cm,clip,width=5.7cm,angle=90]{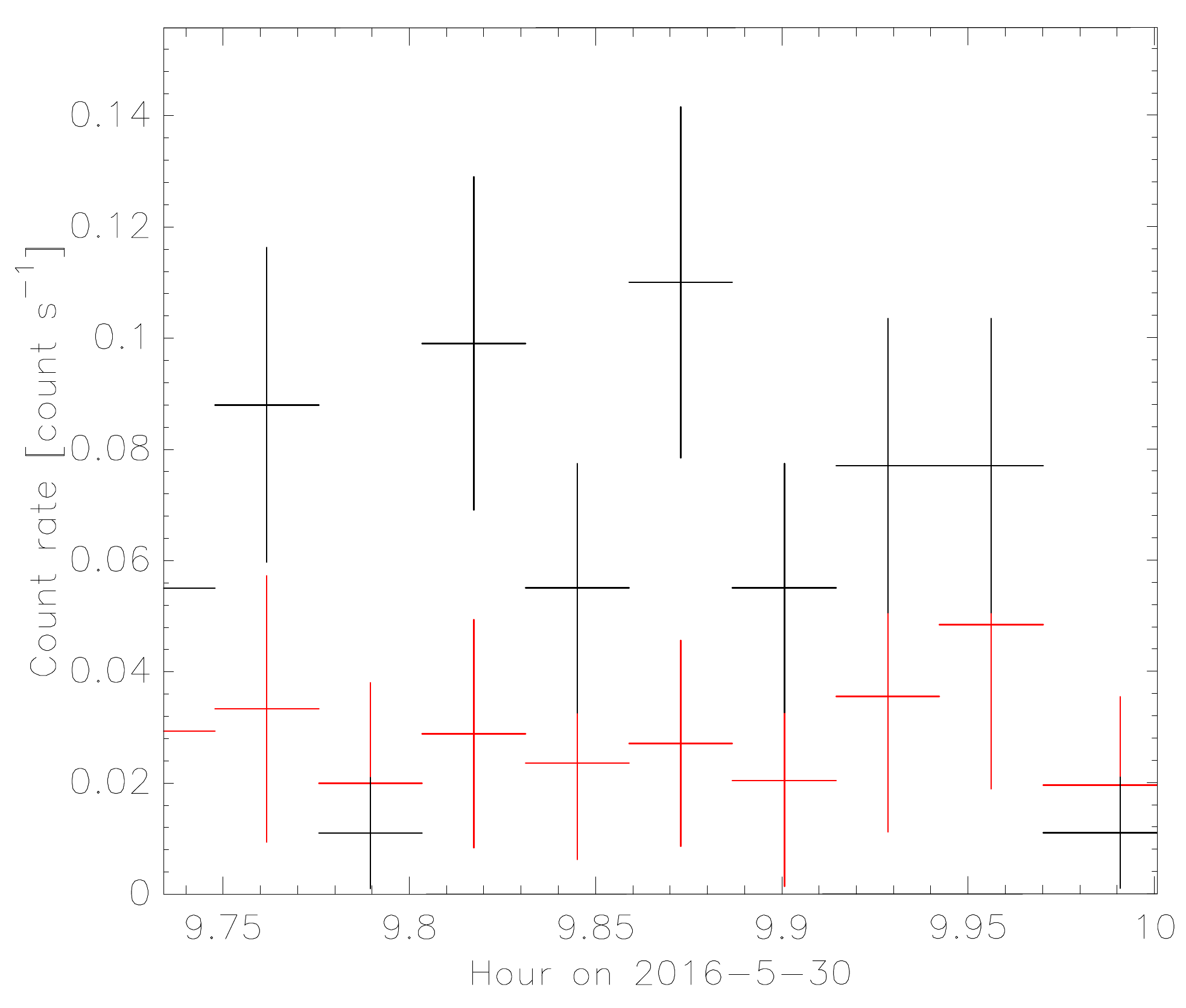}
\caption[The Swift light curve of the 2016 flares]{Light curves with a time bin of 100\,s as observed by \textit{Swift} during the active phase of the two transients.
The black lines show the \sgra{} count rate, while the red lines represent the average light curve of the PSF tail of the two transients at the same distance as \sgra{}.
The vertical bars are the 1$\sigma$ error bars.}
\label{swift_flare_100}
\end{figure*}

During the active phase of the two transients, four observations were tagged as flares (from flare 1 to 4).
To determine whether these increases in flux can be attributed to \sgra{} flares, we first constructed the light curves of these observations with a time bin of 100\,s (Fig.~\ref{swift_flare_100}).
During the two fist observations (flares 1 and 2), the count rate from \sgra{} is clearly above the average count-rate of the PSF tail of the two transients at the distance of \sgra{}.
Moreover, the variation in the count rate of \sgra{} does not follow the rates of the two transients.
These two observations are therefore likely attributable to \sgra{} flares.
However, during the last two observations (flares 3 and 4), the error bars are quite large and the variation in the count rate of \sgra{}  seems to follow the rates of the transients.
These observations can therefore not be unequivocally attributed to \sgra{}.
We therefore excluded them from the study of the flaring rate.

\begin{figure*}[t]
\centering
\includegraphics[trim=0.cm 0.cm 0cm 0cm,clip,width=5.9cm,angle=90]{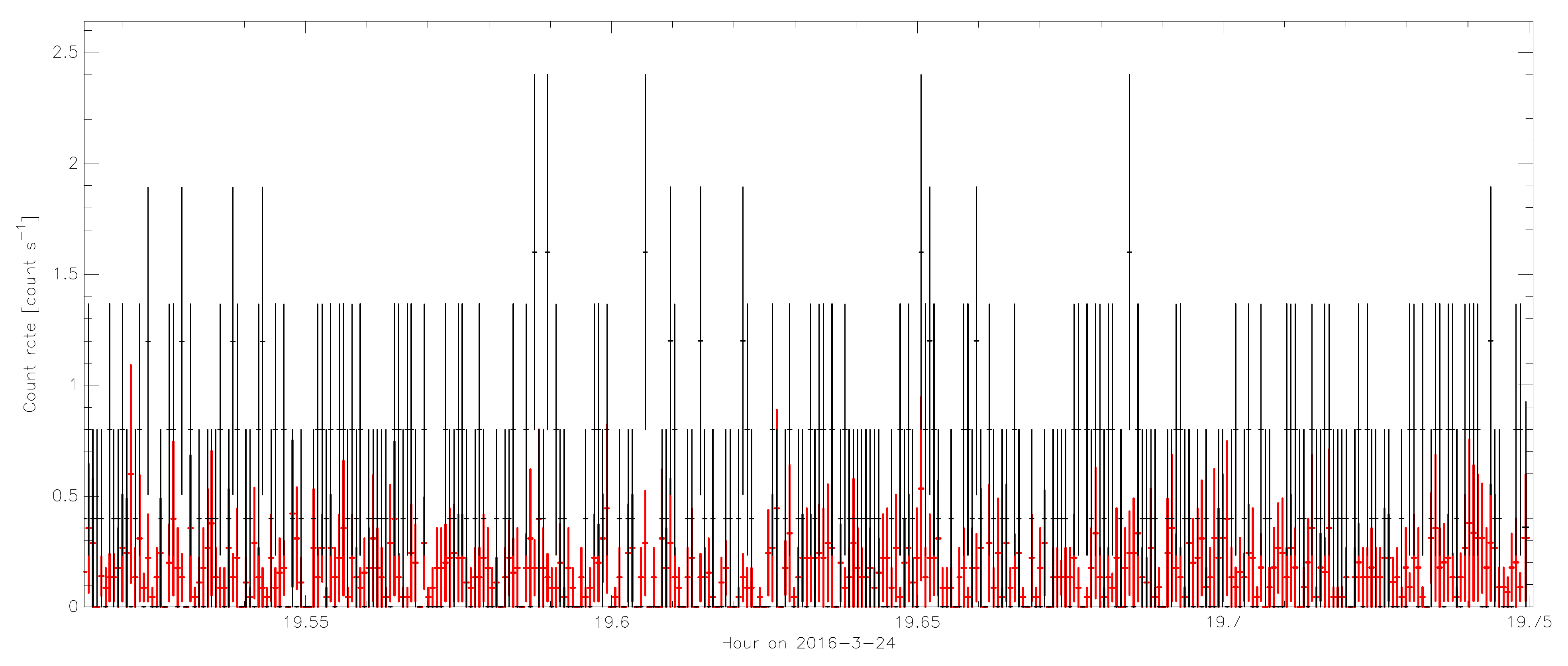}\\
\includegraphics[trim=0.cm 0.cm 0cm 0cm,clip,width=5.9cm,angle=90]{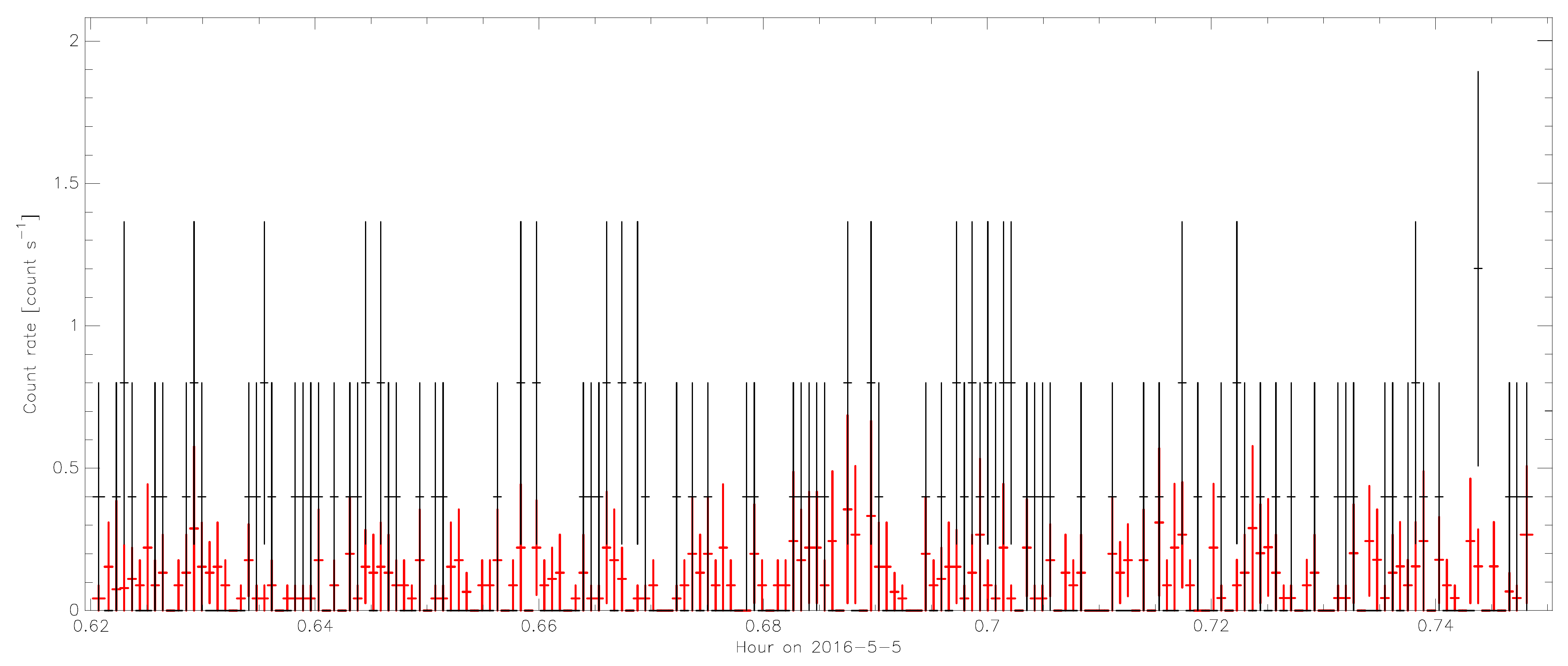}
\caption[The Swift light curve of the 2016 flares]{Light curves with a time bin of 2.5\,s as observed by \textit{Swift} during the first two observations in 2016 that were tagged as flares.
The black lines show the \sgra{} count rate and the red lines represent the average light curve of the PSF tail of the two transients at the same distance as \sgra{}.
The vertical bars are the 1$\sigma$ error bars.}
\label{swift_flare_2}
\end{figure*}

Then, to confirm that the two first observations (flares 1 and 2) are produced by \sgra{}, we also constructed the light curves of these observations using a time bin of 2.5\,s, corresponding to the temporal resolution of the XRT camera (Fig.~\ref{swift_flare_2}).
Magnetars are known to produce single-frame flares that when they are binned on a large time bin produce a shape similar to what is observed during a flare of \sgra{} .
In the light curve of the two observations, the highest count-rates (containing three or four photons in a single frame) are not simultaneous with an increase of flux from the transients.
Flares 1 and 2 can therefore be attributed to \sgra{} without doubt and will therefore be used in the analysis of the flaring rate.

\end{appendix}

\bibliographystyle{aa}
\bibliography{biblio}

\end{document}